\documentclass[aps,twocolumn,pre]{revtex4}
\usepackage{graphicx}
\usepackage{amsfonts}
\usepackage{amssymb}
\usepackage{amsmath}

\def\wfig{4.25cm}
\def\hfig{3.0cm}

\begin{document}

\title{Statics, Dynamics and Manipulations of Bright Matter-Wave Solitons in
Optical Lattices}
\author{P.G. Kevrekidis$^1$, D.J.\ Frantzeskakis$^{2}$, R.\
Carretero-Gonz\'alez$^3$, B.A. Malomed$^4$, G. Herring$^1$ and
A.R. Bishop$^{5}$} \affiliation{$^1$ Department of Mathematics and
Statistics, University of Massachusetts,
Amherst MA 01003-4515, USA \\
$^{2}$ Department of Physics, University of Athens, Panepistimiopolis,
Zografos, Athens 15784, Greece \\
$^3$ Nonlinear Dynamical Systems Group, Department of Mathematics and
Statistics, San Diego State University, San Diego CA, 92182-7720, USA,
http://nlds.sdsu.edu/ \\
$^{4}$ Department of Interdisciplinary Studies, Faculty of Engineering, Tel
Aviv University, Tel Aviv 69978, Israel \\
$^{5}$ Center for Nonlinear Studies and Theoretical Division, Los Alamos
National Laboratory, Los Alamos, NM 87545, USA }

\date{In Press Phys.\ Rev.\ A, 2005}

\begin{abstract}
Motivated by recent experimental achievement in the work with Bose-Einstein
condensates (BECs), we consider bright matter-wave solitons, in the presence
of a parabolic magnetic trap and a spatially periodic optical lattice (OL),
in the attractive BEC. We examine pinned states of the soliton and their
stability by means of perturbation theory. The analytical predictions
are found to be in good agreement with numerical simulations. We then
explore possibilities to use a time-modulated OL as a means of 
stopping and trapping
a moving soliton, and of transferring an initially 
stationary soliton to a prescribed
position by a moving OL. We also study the emission of radiation from the
soliton moving across the combined magnetic trap and OL. We find 
that the soliton moves freely (without radiation) across a weak lattice, 
but suffers strong loss for stronger OLs.
\end{abstract}

\maketitle

%\textwidth 15.5cm

%\email{kevrekid@math.umass.edu}

\section{Introduction\label{sec1}}

The recent progress in experimental and theoretical studies of Bose-Einstein
condensates (BECs) \cite{reviews} has led to an increase of interest in
matter-wave (MW)\ solitons. One-dimensional (1D) dark \cite{expd} and bright
\cite{expb} solitons have been observed in experiments with repulsive and
attractive BECs, respectively. Very recently, bright solitons of the gap
type, predicted in repulsive condensates \cite{md}, have been created in the
experiment \cite{Oberthaler}. Theoretical predictions concerning a
possibility of the existence of stable multi-dimensional solitons supported
by a full \cite{md,BBB1} or low-dimensional \cite{lowD} optical lattice (OL)
have also been reported. The OL is created as a standing-wave interference
pattern between mutually coherent laser beams \cite{g2,g3,g4,g5,g6,g7}.

The study of the MW solitons, apart from being a fundamentally interesting
topic, may have important applications. In particular, a soliton may be
transferred and manipulated similarly to what has been recently
demonstrated, experimentally and theoretically, for BECs in magnetic
waveguides \cite{aw} and atom chips \cite{ac}. More generally, the
similarity between bosonic MWs and light waves suggests that numerous
results known for optical solitons \cite{ka}, along with the possibility of
manipulation of atomic states (by means of resonant electromagnetic waves
governing transitions between different states), may have impact on the
rapidly evolving field of quantum atom optics (see, e.g., Ref. \cite{mo}).

A context where the dynamics of MW solitons is particularly interesting is
that of BECs trapped in a periodic potential induced by the above-mentioned
OLs. The possibility to control the OL has led to the realization of many
interesting phenomena, including Bloch oscillations \cite{g4,g8},
Landau-Zener tunneling \cite{g2} (in the presence of an additional linear
external potential), and classical \cite{Catall} and quantum \cite{g7}
superfluid-insulator transitions. A large amount of theoretical work has
been already done for nonlinear MWs trapped in OLs (see Refs.\ \cite{pd,bk}
for recent reviews).

The objective of this work is to systematically study the statics and
dynamics of one-dimensional (1D) bright MW solitons confined in the
combination of the parabolic magnetic trap (MT) and OL. Additionally, we
examine the possibility to control the motion of the soliton by means of a
\textit{time-dependent} OL potential (the latter is available for the
experiment). In particular, we will show that, in the case when the OL
period is comparable to the characteristic spatial width of the soliton, it
is possible to: (a) snare and immobilize an originally moving soliton in a
local potential well, by adiabatically switching the OL on, and (b) grasp
and drag an initially stationary soliton by a slowly moving OL, delivering
it to a desired location. Note that bright MW solitons may travel long
distances in the real experiment, up to several millimeters \cite{expb}, and
are truly robust objects, being themselves coherent condensates. Thus,
the manipulation of bright MW solitons is a very relevant issue for the
physics of BECs.

The paper is organized as follows. In Sec.\ \ref{sec2}, we introduce the
model and present analytical results. In Sec.\ \ref{sec3}, we numerically
investigate static and dynamical properties of the solitons, and study
possibilities to manipulate them as outlined above. The results of the work
are summarized in Sec. \ref{sec4}.

\section{The model and its analytical consideration\label{sec2}}

%Setup and Analytical Results

The Gross-Pitaevskii equation (GPE), which governs the evolution
of the single-atom wave function in the mean-field approximation,
takes its fundamental form in the 3D case. A number of works
analyze its reduction to an effective 1D equation in the case of
strongly elongated cigar-shaped BECs
\cite{theory,Isaac,Salasnich}. In particular, the derivation in
Ref. \cite{Isaac} assumed that the potential energy is much larger
than the transverse kinetic energy. A general conclusion is that
the effective equation reduces to the straightforward 1D version
of the GPE. In the normalized form, it reads \cite{pd}
\begin{equation}
iu_{t}=-\frac{1}{2}u_{xx}+g|u|^{2}u+V(x)u,  \label{meq1}
\end{equation}
where $u(x,t)$ is the 1D mean-field wave function (although a different
version of the 1D GPE, with a non-polynomial nonlinearity, is known
too \cite{Salasnich}). The combination of the MT and OL potential
corresponds to
\begin{equation}
V(x)=\frac{1}{2}\Omega ^{2}x^{2}+V_{0}\sin ^{2}(kx).  
\label{meq2}
\end{equation}
In Eq.\ (\ref{meq1}), $x$ is measured in units of the fluid healing length $\xi=\hbar/\sqrt{n_{0} g_{1D}m}$, where $n_{0}$ is the peak density and $g_{1D} \equiv g_{3D}/(2\pi l_{\perp}^{2})$ is the effective 1D interaction strength, obtained upon integrating the 3D interaction strength $g_{3D}=4\pi \hbar^{2}a/m$ in the transverse directions ($a$ is the scattering length, $m$ the atomic mass, and 
$l_{\perp}=\sqrt{\hbar / m \omega_{\perp}}$ is the transverse harmonic oscillator length, with $\omega_{\perp}$ being the transverse-confinement frequency). Additionally, $t$ is measured in units of $\xi/c$ (where $c=\sqrt{n_{0}g_{1D}/m}$ is the Bogoliubov speed of sound), the atomic density is rescaled by the peak density $n_{0}$, and energy is measured in units of the chemical potential of the 
system $\mu=g_{1D}n_{0}$. Accordingly, the dimensionless parameter $\Omega \equiv \hbar \omega _{x}/g_{1D}n_{0}$ ($\omega _{x}$ is the confining frequency in the axial direction) determines the magnetic trap strength, $V_{0}$ is the OL strength, while  
$k$ is the wavenumber of the OL; the latter, can be controlled by varying the angle $\theta $ between the counter-propagating laser
beams, so that $\lambda \equiv 2\pi /k=\left( \lambda_{\mathrm{laser}}/2\right) \,\sin (\theta /2)$ \cite{arimondo}. Finally, $g=\pm 1$ is the renormalized nonlinear coefficient, which is positive (negative) for a repulsive (attractive) condensate. As we are interested in the ordinary bright MW solitons, which exist in case of attraction, we will fix $g=-1$.

Without the external potential ($\Omega =V_{0}=0$), Eq.\ (\ref{meq1})
supports bright soliton solutions of the form
\begin{equation}
u_{s}(x-x_{0})=\eta ~\mathrm{sech}\left[ \eta (x-x_{0})\right] \exp \left(
\frac{1}{2}i\eta ^{2}t\right) ,  \label{meq3}
\end{equation}where $\eta $ is the soliton's amplitude, and $x_{0}$ is the coordinate of
its center. Moving solitons can be generated from the zero-velocity one by
a Galilean boost.

In the presence of the external potential, the first issue is to identify
stationary positions for the soliton. This issue can be addressed, using an
effective potential for the soliton's central coordinate (see, e.g., Refs.\
\cite{RMP} and \cite{sb}), which is defined by the integral
\begin{equation}
V_{\mathrm{eff}}(x_{0})=\int_{-\infty }^{+\infty }V(x)|u_{s}(x-x_{0})|^{2}dx.
\label{meq4}
\end{equation}
Stationary positions of the soliton correspond to local extrema of the
effective potential (\ref{meq4}). This well-known heuristic result can be
rigorously substantiated by means of the Lyapunov-Schmidt theory applied to
the underlying nonlinear Schr{\"{o}}dinger equation \cite{kapitula}. The
effective potential corresponding to the external potential (\ref{meq2}),
acting on the stationary soliton (\ref{meq3}), can be easily evaluated:
\begin{equation}
V_{\mathrm{eff}}(x_{0})=\eta \Omega ^{2}x_{0}^{2}-\pi V_{0}k\cos
(2kx_{0})\mathrm{csch}\left( \frac{k\pi }{\eta }\right) .
\label{meq5}
\end{equation}
Notice that, depending on values of the parameters, this potential may have
a single extremum at $x_{0}=0$, or multiple ones.

The stability of the soliton resting at a stationary position can
also be analyzed in terms of the effective potential (\ref{meq4}):
the position is stable if it corresponds to a potential minimum.
This well-known result can be rigorously derived using the theory
of Ref.\ \cite{grillakis} and reformulated in Ref.\
\cite{kapkevsan} (see also Refs.\ \cite{pelinovsky} and
\cite{oh}). In particular, the curvature of the potential at the
stationary position determines the key linearization eigenvalue
$\lambda $, that may cause an instability, bifurcating through the
origin of the corresponding spectral plane (this feature is
revealed by the heuristic \cite{RMP} and rigorous \cite{kapkevsan}
analysis). The eigenvalue is easily found to be
\begin{equation}
\lambda ^{2}=-\eta ^{-1/2}V_{\mathrm{eff}}^{\prime \prime }(x_{0}),
\label{meq6}
\end{equation}confirming that minima and maxima of the effective potential (\ref{meq4})
give rise, respectively, to stable ($\lambda ^{2}<0$) and unstable ($\lambda
^{2}>0$) equilibria.

We note in passing (this will be important in what follows) that
the minima of the effective potential (\ref{meq4}) \textit{differ}
from the minima of the external potential $V(x)$ trapping the
atoms. For instance, for $\eta =\sqrt{2}$, $V_{0}=0.25$ and
$\Omega =0.1$, the first three minima of $V(x)$ (apart from the
one at $x=0$) are located at the points $x=3.0789,6.1587,9.2356$,
while the minima of $V_{\mathrm{eff}}$ are found at
$x_{0}=3.0166,6.0247,9.0089$.

We now turn to numerical results, aiming to examine the validity of the
theoretical predictions, as well as to perform dynamical experiments using
the OL to guide the soliton motion.

\section{Numerical results\label{sec3}}

\subsection{Stability of the solitons}

We begin the numerical part by examining the steady-state soliton solutions
and their stability in the context of Eq.\ (\ref{meq1}). Such solutions are
sought for in the form $u(x,t)=\exp (i\Lambda t)w(x)$, which results in the
stationary equation,
\begin{equation}
\Lambda w=(1/2)w_{xx}+w^{3}-V(x)w.  \label{meq7}
\end{equation}To examine the linear stability of the solitons, we take a perturbed
solution as
\begin{equation}
u(x)=e^{i\Lambda t}\left[ w+\epsilon \left( a(x)e^{-i\omega
t}+b(x)e^{i\omega ^{\ast }t}\right) \right] ,  \label{meq8}
\end{equation}where $\epsilon $ and $\omega $ are an infinitesimal amplitude and
(generally speaking, complex) eigenfrequency of the perturbation, and
linearize Eq.\ (\ref{meq1}).

Equations (\ref{meq1}) and (\ref{meq2}), with $g=-1$ and arbitrary
coefficients $V_{0},\Omega $ and $k$, possess a scaling
invariance, which allows us to fix $\Lambda =1$ (hence $\eta
=\sqrt{2}$). It should be noted that, in the absence of the MT
($\Omega=0$), the soliton's frequency should be chosen so that it
belongs to a bandgap in the spectrum of the linearized
Eq.\ (\ref{meq1}) with the periodic potential (\ref{meq2}), to avoid
resonance with linear Bloch waves. However, the MT potential with
finite $\Omega$ makes this condition irrelevant. In principle, it
might be interesting to investigate how the increase of $\Omega$
from zero gradually lifts the condition of the resonance
avoidance, but this more formal issue is left beyond the scope of
the present work.

To estimate actual physical quantities corresponding to the above normalized 
values of the parameters, we consider a cigar-shaped $^{7}$Li condensate 
containing $N \simeq 10^{3}$ atoms in a trap with $\omega _{x}=2\pi \times 25$ Hz and 
$\omega _{\perp }= 70 \omega _{x}$. Then, for a 1D peak density $n_{0}=10^8$ ${\rm m}^{-1}$, 
the parameter $\Omega$ in Eq.\ (\ref{meq2}) assumes the value $\Omega =0.1$, while the time and
space units correspond to $0.3$ ms and $1.64 \mu$m, respectively. These units
remain valid for other values of $\Omega $, as one may vary $\omega _{\perp }
$ and change $\omega _{x}$ accordingly; in this case, other quantities, such
as $N$, also change.
%, which corresponds to the real experiments \cite{expb}.

%%%%%%%%%%%%%%%%%%%%%%%%%%%%%%%%%%%%%%%%%%%%%%%%%%%%%%%%%%%%%%%%%%%%%%%%%%%
\begin{figure}[t]
\centerline{
\includegraphics[width=\wfig,height=\hfig,angle=0,clip]{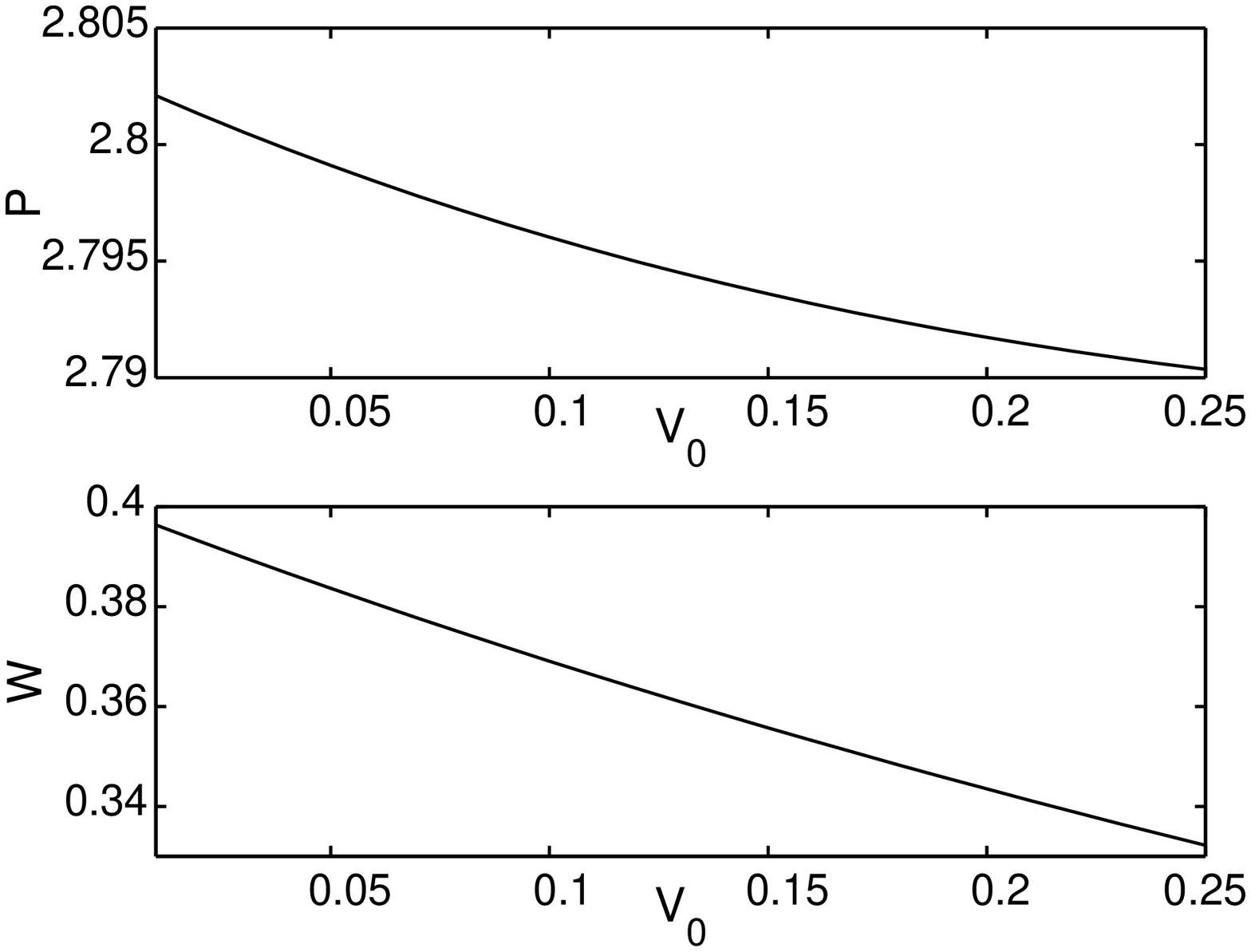}
\includegraphics[width=\wfig,height=\hfig,angle=0,clip]{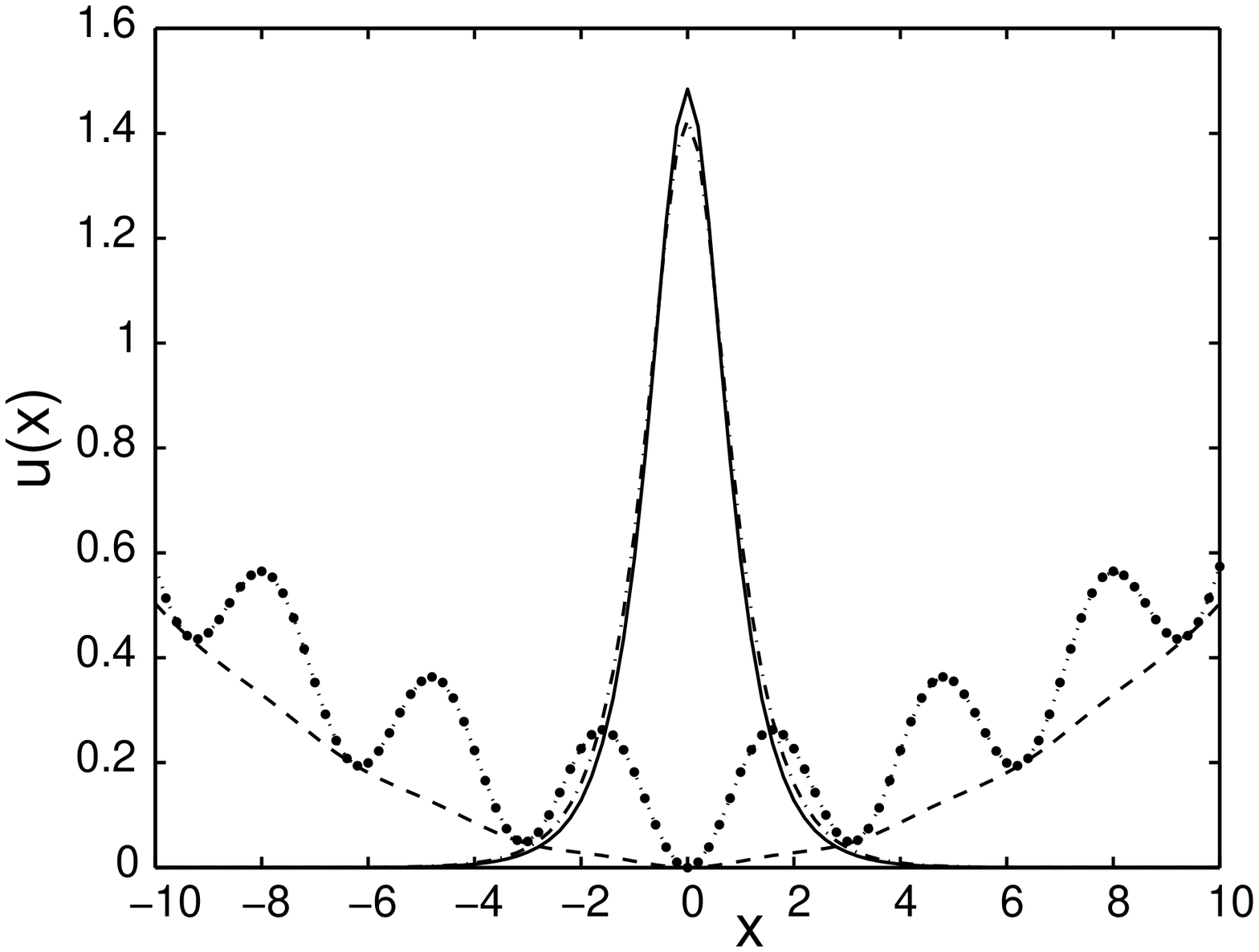}
}
\centerline{
\includegraphics[width=\wfig,height=\hfig,angle=0,clip]{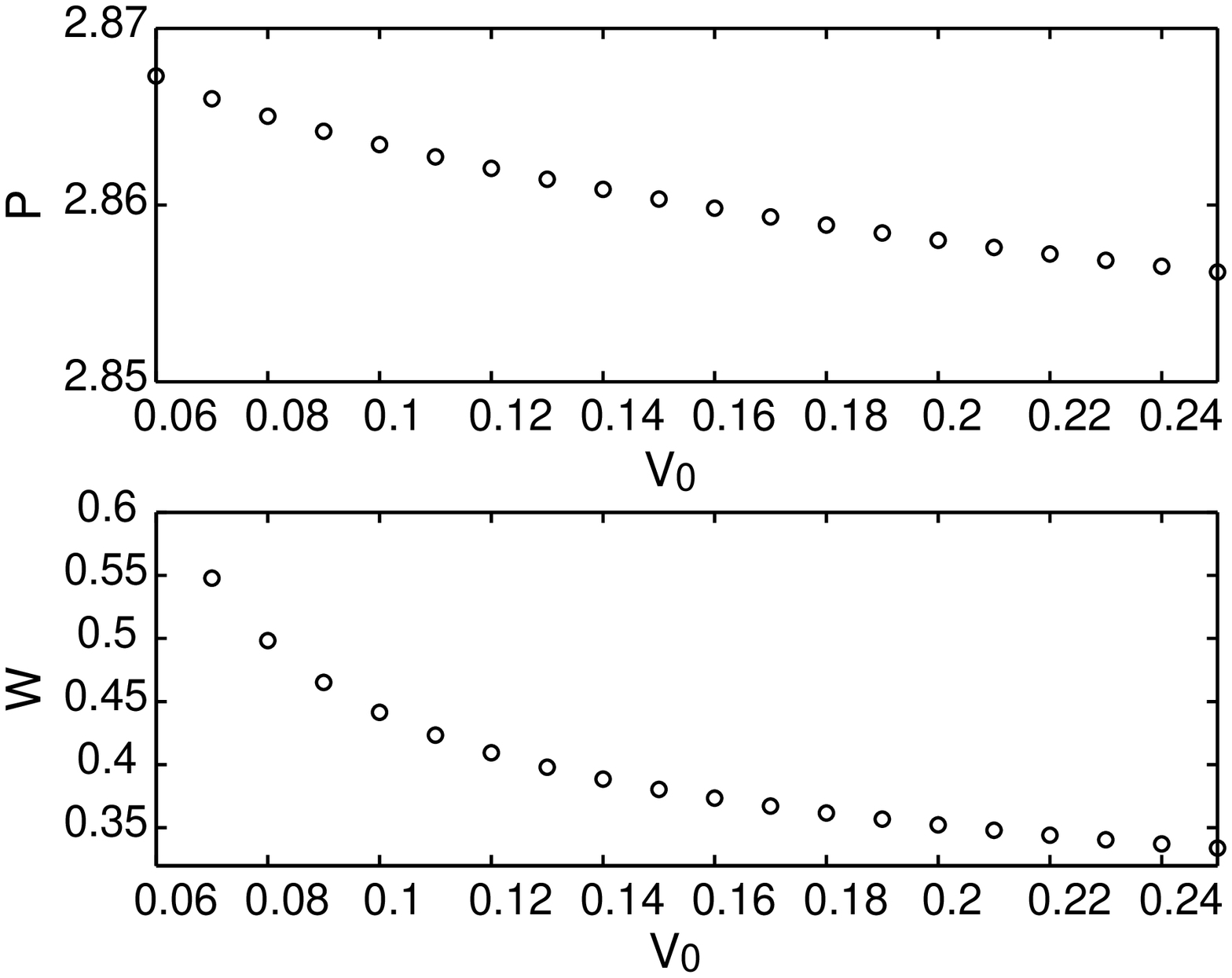}
\includegraphics[width=\wfig,height=\hfig,angle=0,clip]{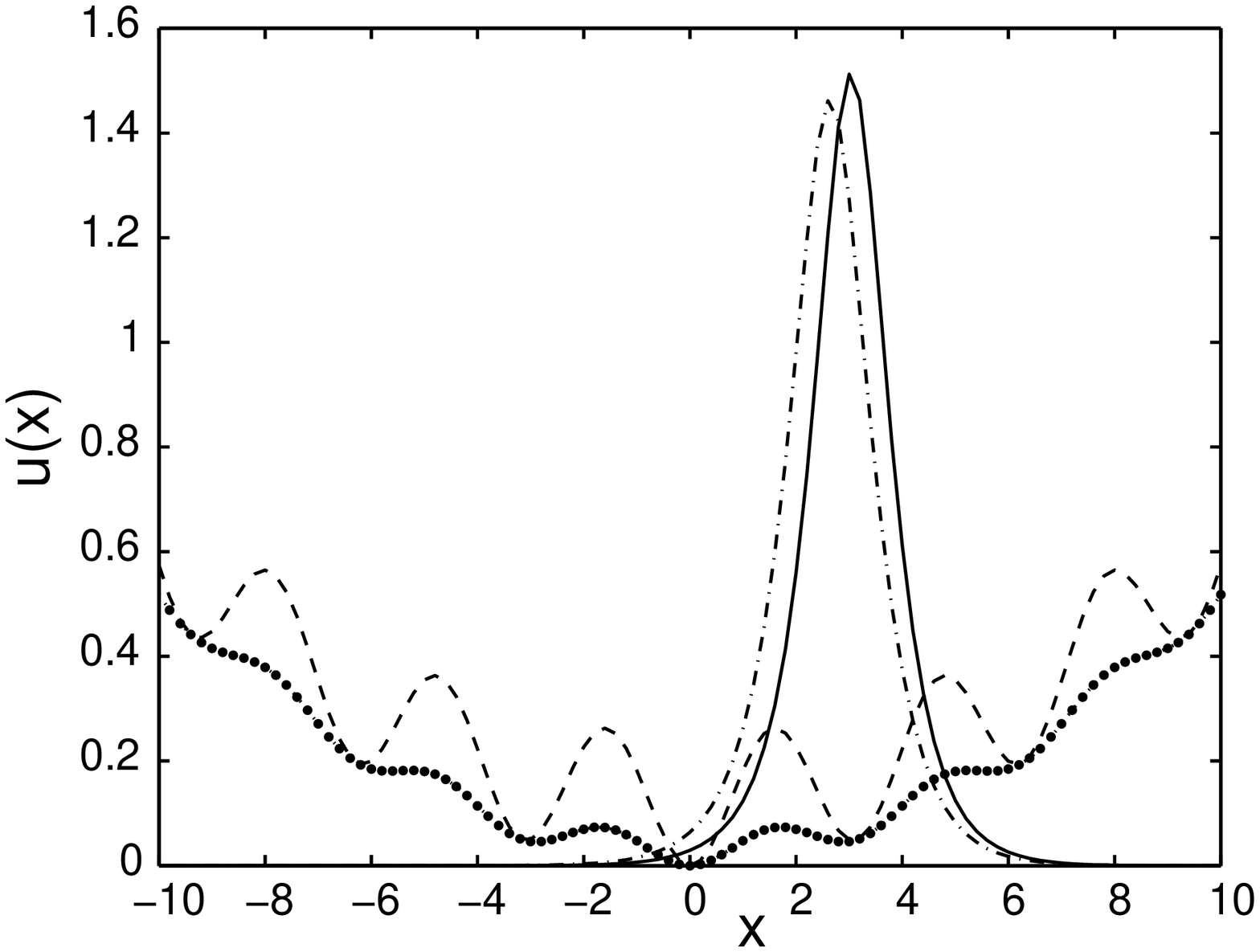}
}
\centerline{
\includegraphics[width=\wfig,height=\hfig,angle=0,clip]{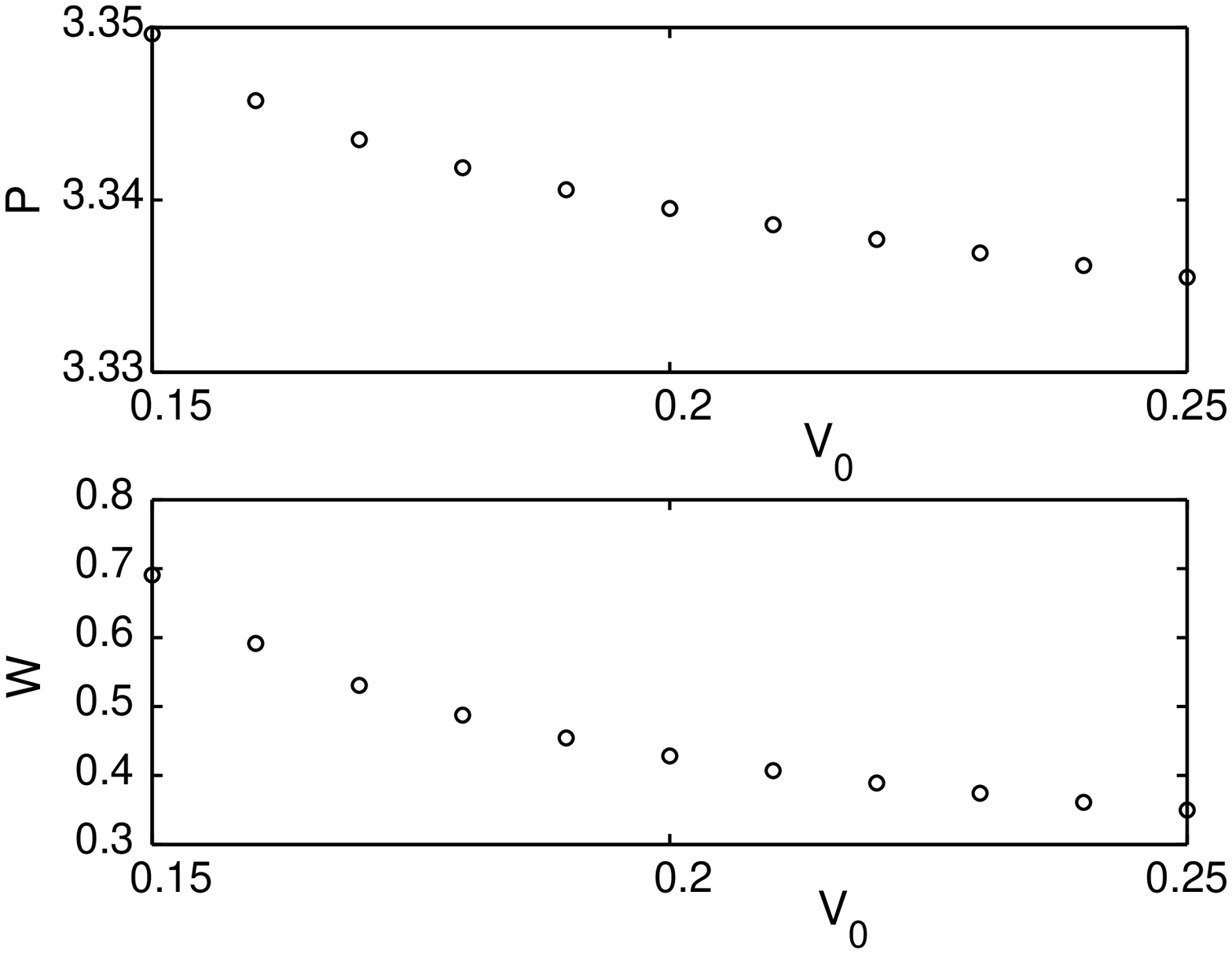}
\includegraphics[width=\wfig,height=\hfig,angle=0,clip]{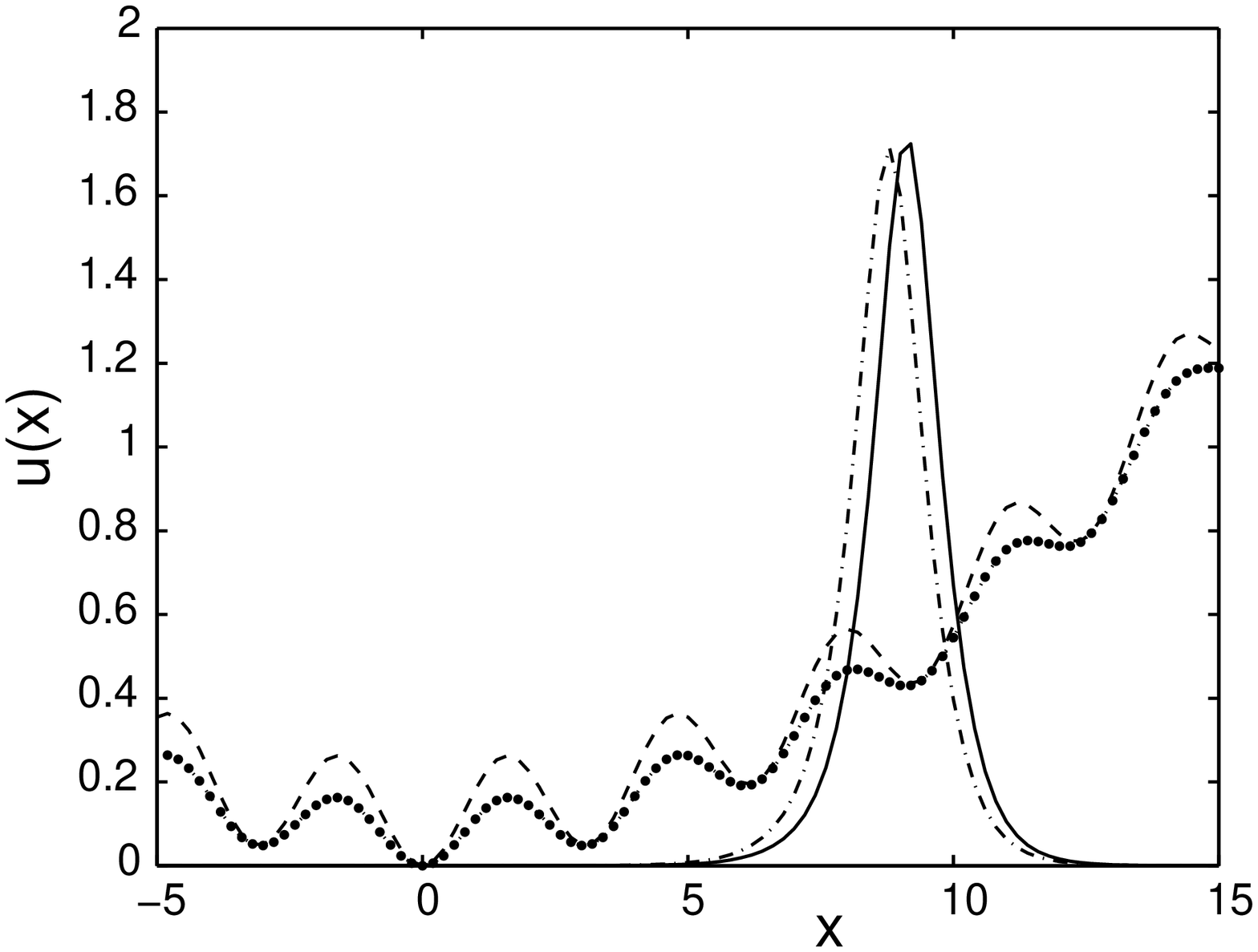}
} \caption{For each of the three sets of the pictures, the left
panel shows the continuation of the soliton branch to values near
$V_{0}^{(\mathrm{cr})}$, at which it disappears (for soliton
solutions trapped at different wells). The right panel shows the
solution at the initial and final points of the continuation (and
the corresponding potentials). The left panels show the norm of
the soliton solution (proportional to the number of atoms in the
condensate), $P=\protect\int_{-\infty }^{+\infty }|u(x)|^{2}dx$,
and its squared width, $W=P^{-1}\protect\int_{-\infty }^{+\infty
}x^{2}|u(x)|^{2}dx$, as a function of the OL strength $V_{0}$. The
top set of the panels pertains to the zeroth-well solution (the
soliton pinned in the central potential well); the solution in the
right panel is shown by the solid line for $V_{0}=0.25$, and by
the dash-dotted line for $V_{0}=0$. The corresponding potential is
shown by the dotted line for $V_{0}=0.25$, and by the dashed line
for $V_{0}=0$. Similarly, in the middle set, the solid line (and
the dashed one for the potential) correspond to $V_{0}=0.25$, and
the dashed-dotted line, together with the dotted one for the
potential, correspond to $V_{0}=0.06$ for the first-well solution
[notice that this branch terminates at $V_{0}\approx 0.045$].
Finally, in the bottom set of the panels, the solid line (and the
dashed one for the potential) again correspond to $V_{0}=0.25$,
while the dashed-dotted line (and the dotted one for the
potential) correspond to $V_{0}=0.15$ for the third-well solution
[this branch terminates at $V_{0}\approx 0.1425$].} \label{Fig1}
\end{figure}
%%%%%%%%%%%%%%%%%%%%%%%%%%%%%%%%%%%%%%%%%%%%%%%%%%%%%%%%%%%%%%%%%%%%%%%%%%%

Figure\ \ref{Fig1} summarizes our numerical findings for the
stability problem. As expected, the (\textit{zeroth-well})
solution for the soliton pinned at $x_{0}=0$ exists and it is
stable for all values of the potential's parameters. We have
typically chosen to fix $\Omega =0.1$ and $k=1$ (i.e., $\lambda
=2\pi $) and vary $V_{0}$; however, it has been checked that the
results presented below adequately represent the phenomenology for
other values of $(\Omega ,k)$ as well.

The next (\textit{first-well}) solution, corresponding to the
potential minimum closest to $x_{0}=0$, exists for values of the
MT strength $V_{0}$ smaller than a critical one
$V_{0}^{(\mathrm{cr})}$. Within the accuracy of $0.0025$, we have
found its numerical value to be
$V_{0}^{(\mathrm{cr})}|_{\mathrm{num}}=0.045$, in very good
agreement with the prediction following from the analytical
approximation (\ref{meq5}) for the effective potential, which
shows that the corresponding potential minimum disappears, merging
with a maximum, at $V_{0}^{(\mathrm{cr})}|_{\mathrm{anal}}\approx
0.048$. The corresponding pinned-soliton solution is indeed stable
prior to its disappearance, in agreement with the analytical
prediction based on Eq.\ (\ref{meq6}).

Similarly, the subsequent (\textit{second-well}) solution,
associated with the next potential minimum (if it exists), is
found to disappear (for the same parameters) at
$V_{0}^{(\mathrm{cr})}|_{\mathrm{num}}=0.1\pm 0.0025$, while the
analytical approximation (\ref{meq5}) yields
$V_{0}^{(\mathrm{cr})}|_{\mathrm{anal}}\approx 0.112$. Finally, a
similar result was obtained for the third-well solution:
$V_{0}^{(\mathrm{cr})}|_{\mathrm{num}}=0.1425\pm 0.0025$, and
$V_{0}^{(\mathrm{cr})}|_{\mathrm{anal}}\approx 0.176 $.

It is quite natural that the discrepancy between the theoretical and the
numerical results increases for the higher-well solutions, given that the
numerically exact profile of the pinned soliton gets more distorted under
the action of the MT. Notice, for example, the difference in the amplitude
between the soliton in the top panel and in the one in the bottom panel in
Fig.\ \ref{Fig1}, which clearly illustrates this effect.

\subsection{Soliton dynamics and manipulations}

Having addressed the existence and stability of the solitons, we
now proceed to study their possible dynamical
manipulation by means of the OL. First, we examine the
possibility to trap a soliton using the secondary minima in the OL
potential. In particular, it is well known that, in the absence of
the OL, the soliton in the magnetic trap, when displaced from the
center, $x_{0}=0$, executes harmonic oscillations with the
frequency $\Omega $, as a consequence of the Ehrenfest theorem
(alias the Kohn's theorem \cite{kohn}, which states that the
motion of the center of mass of a cloud of particles trapped in a
parabolic potential decouples from the internal excitations). This
result can also be obtained using the variational approximation
\cite{st} and, more generally, is one of the results obtained from
the moment equations for the condensate in the parabolic potential
\cite{victor}.

%%%%%%%%%%%%%%%%%%%%%%%%%%%%%%%%%%%%%%%%%%%%%%%%%%%%%%%%%%%%%%%%%%%%%%%%%%%%%
\begin{figure}[th]
\centerline{
\includegraphics[width=\wfig,height=\hfig,angle=0,clip]{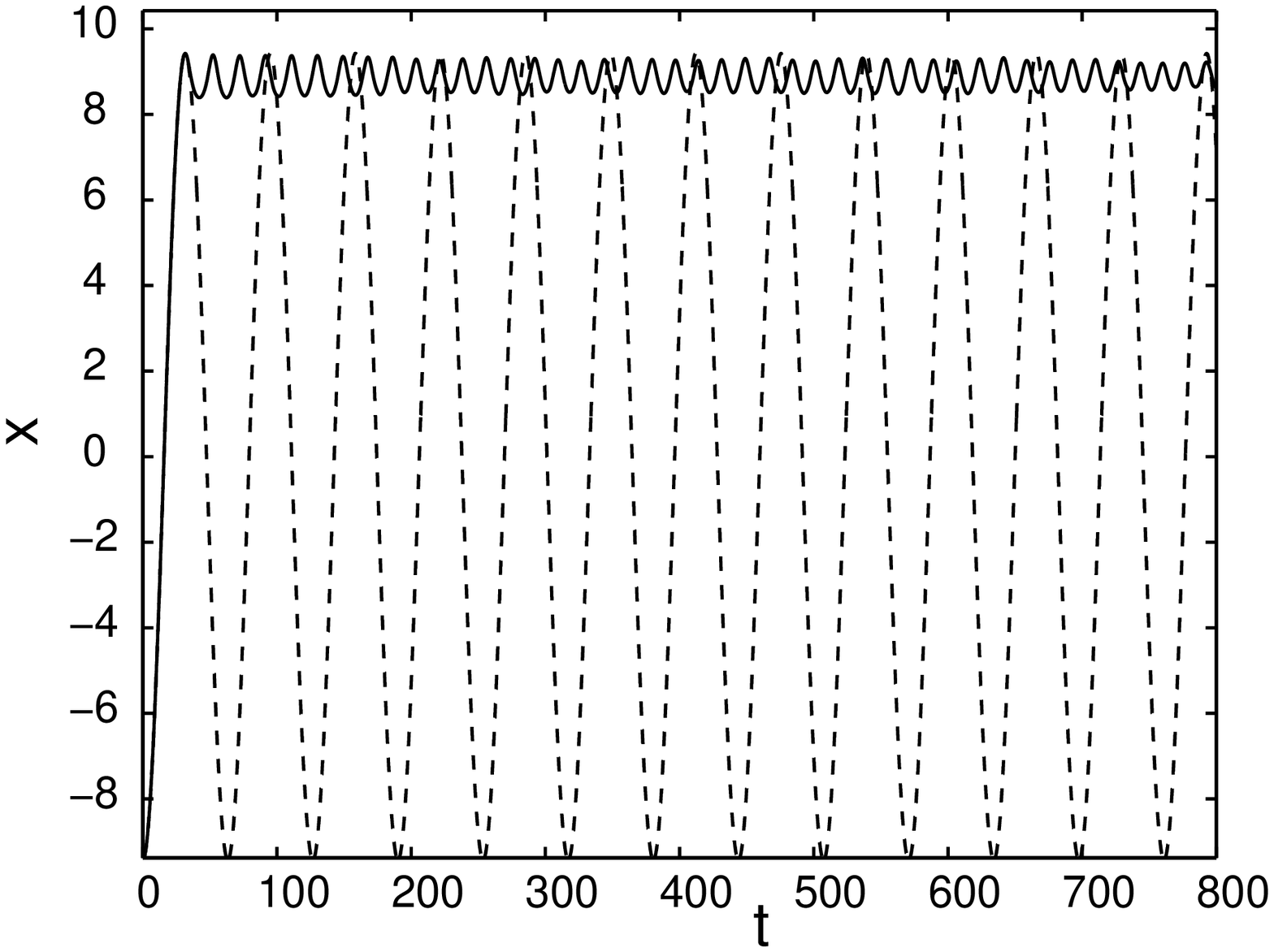}
\includegraphics[width=\wfig,height=\hfig,angle=0,clip]{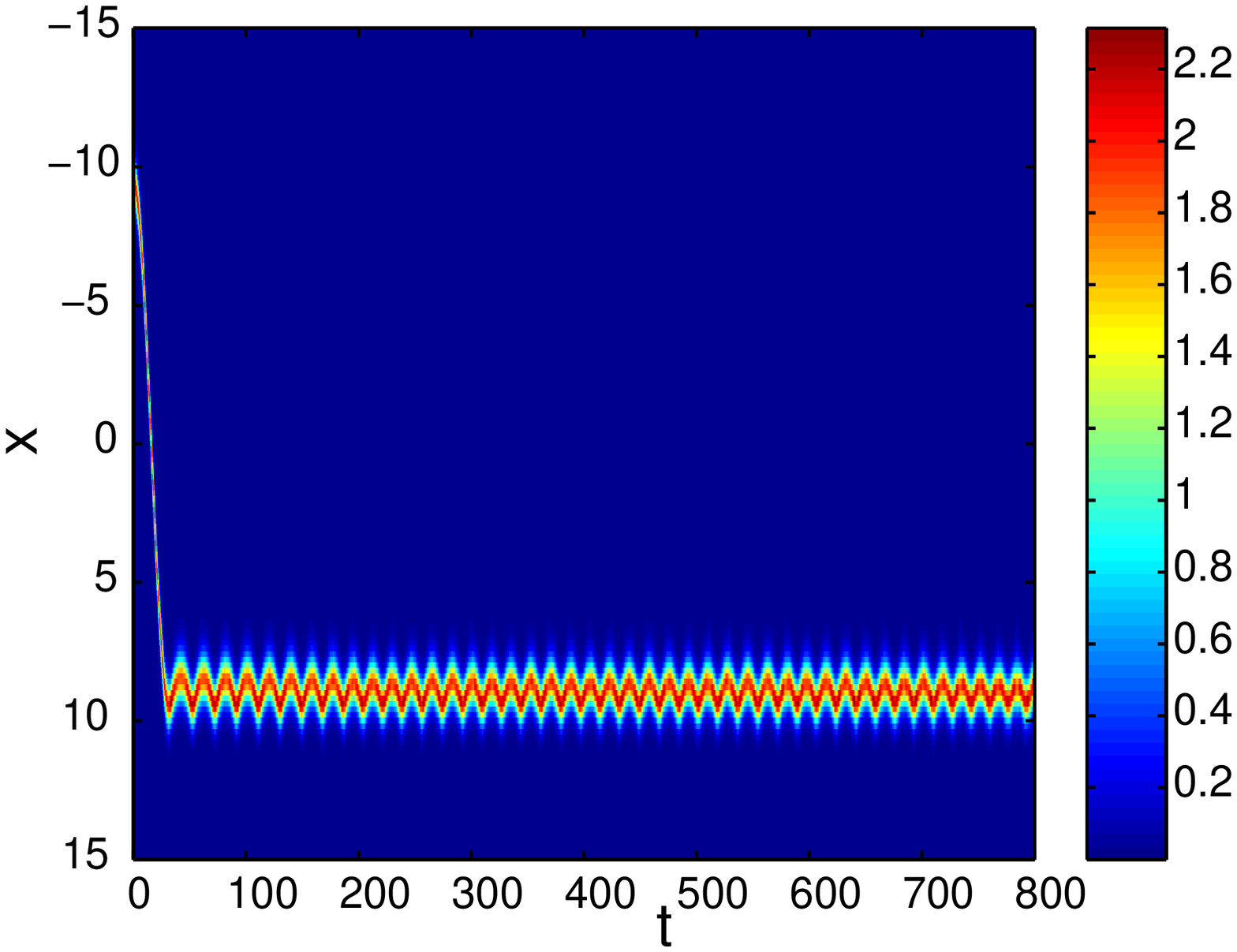}
}
\caption{(color online) An example of snaring the originally moving soliton
using the optical lattice. The left panel shows the motion of the soliton's
center of mass. The dashed line shows the situation without the OL (but in
the presence of the magnetic trap). If we turn on the OL potential, as the
soliton arrives at the turning point of its trajectory, it gets captured by
the secondary minimum of the full potential, created in a vicinity of this
point. The right panel shows the same, but through the space-time contour
plots of the local density, $|u(x,t)|^{2}$.}
\label{Fig1a}
\end{figure}

\begin{figure}[t]
\centerline{
\includegraphics[width=\wfig,height=\hfig,angle=0,clip]{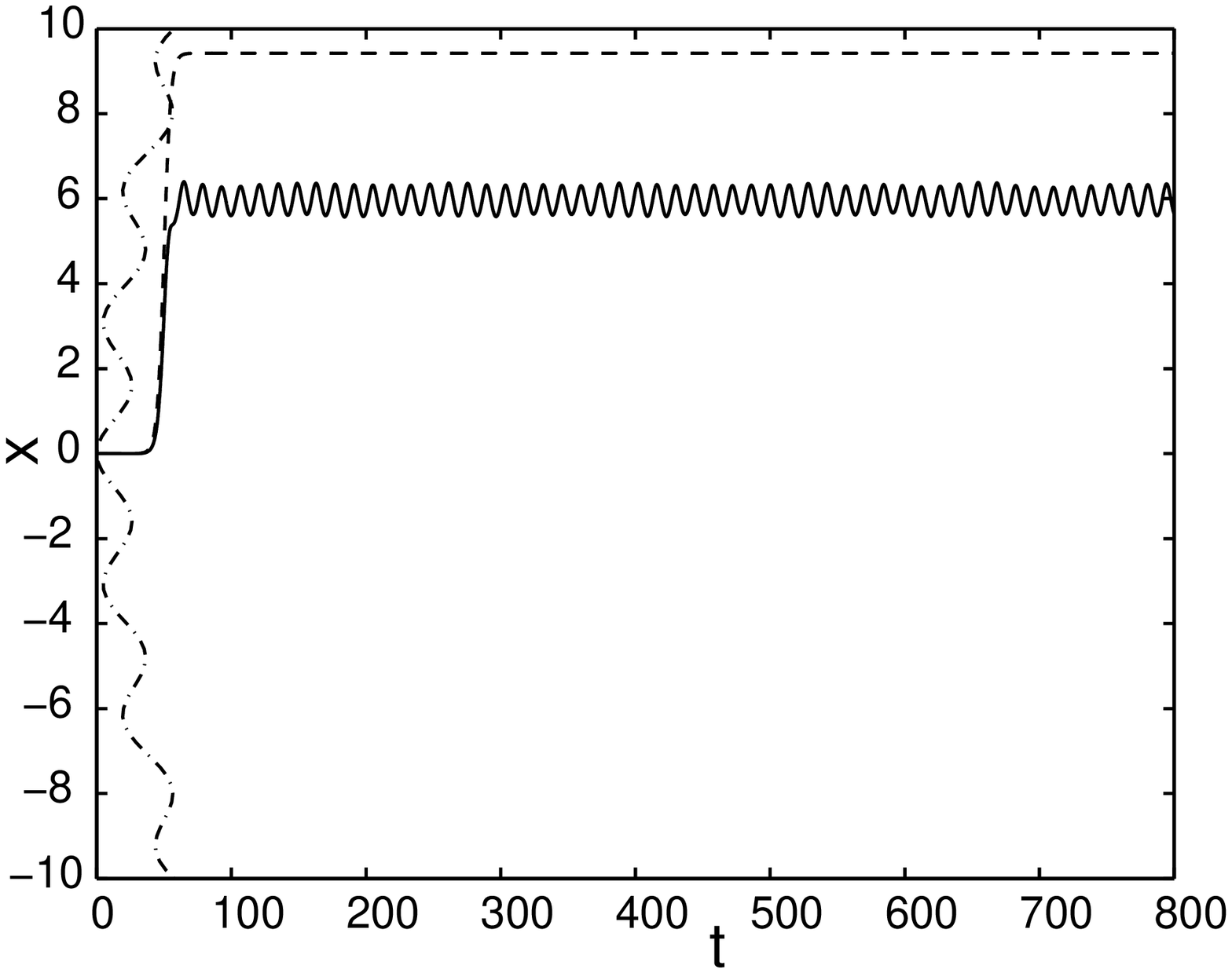}
\includegraphics[width=\wfig,height=\hfig,angle=0,clip]{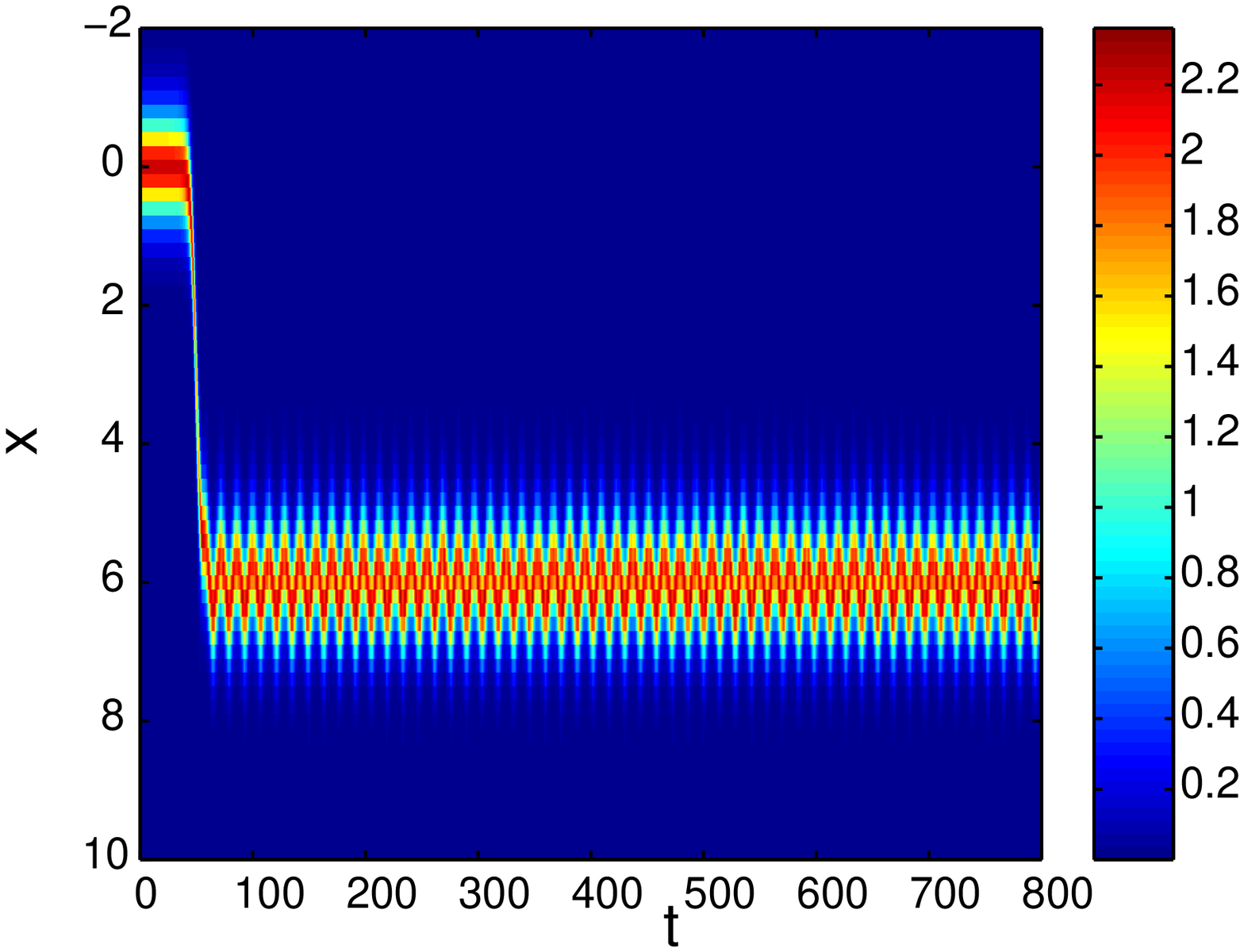}
}
\centerline{
\includegraphics[width=\wfig,height=\hfig,angle=0,clip]{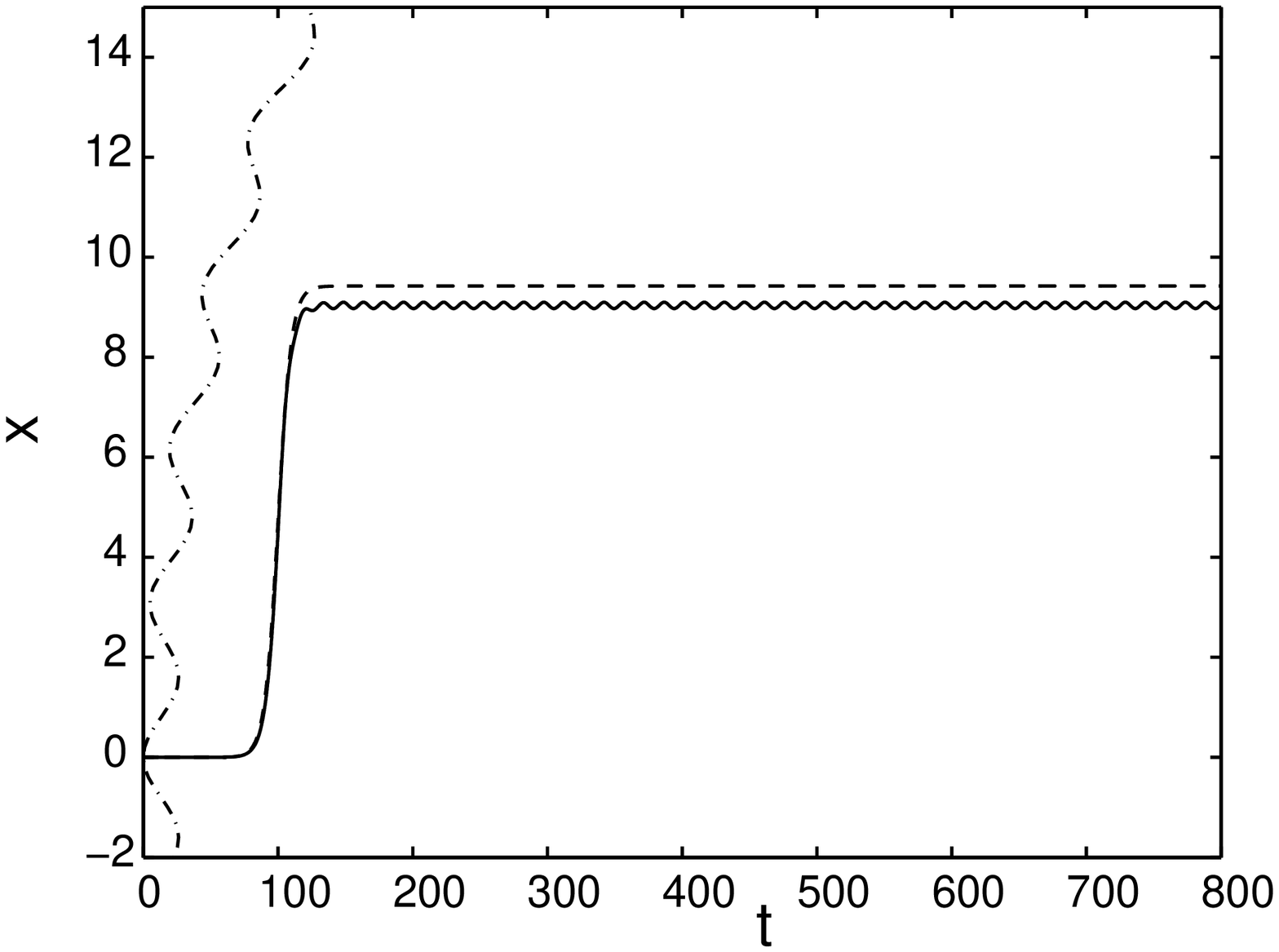}
\includegraphics[width=\wfig,height=\hfig,angle=0,clip]{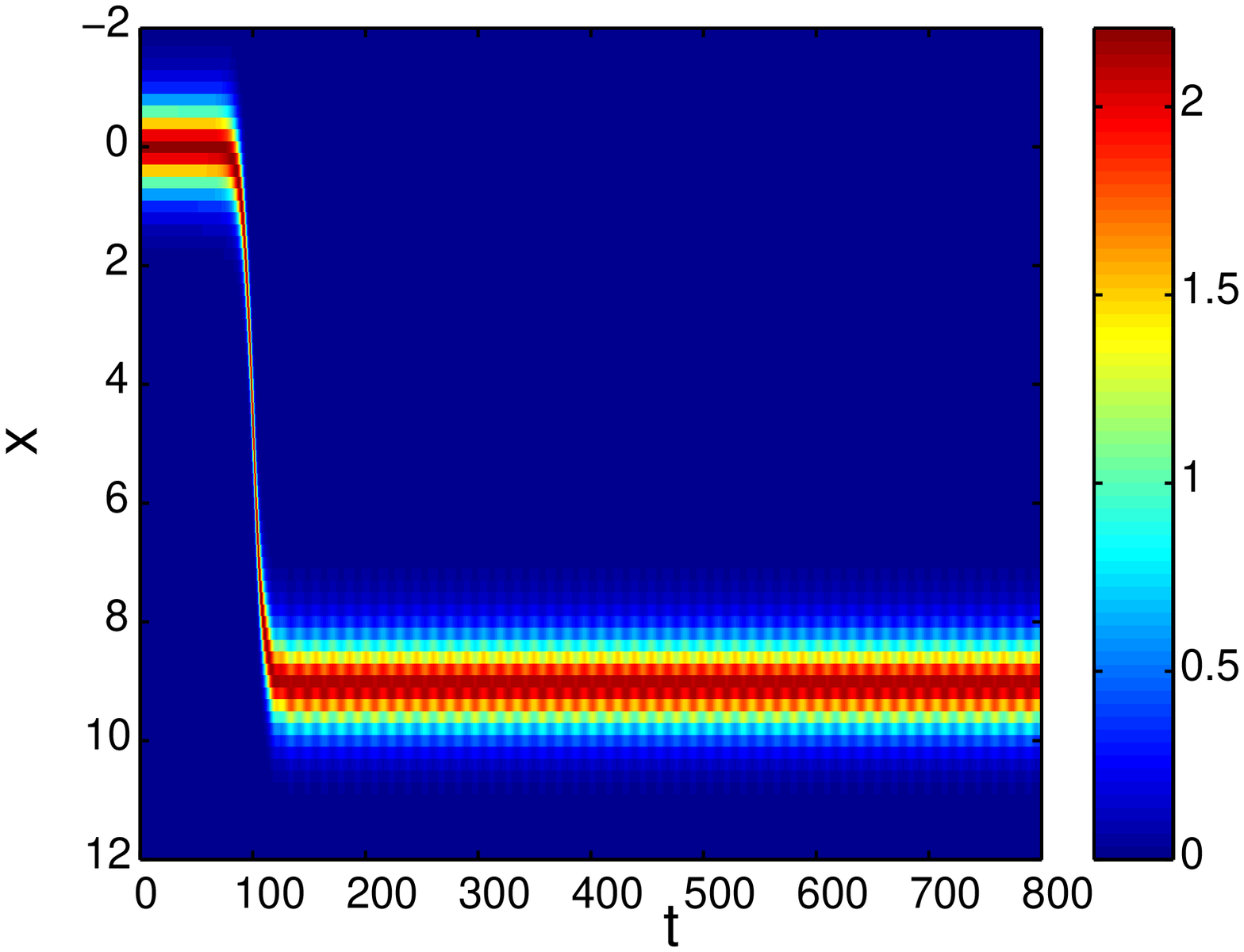}
}
\centerline{
\includegraphics[width=\wfig,height=\hfig,angle=0,clip]{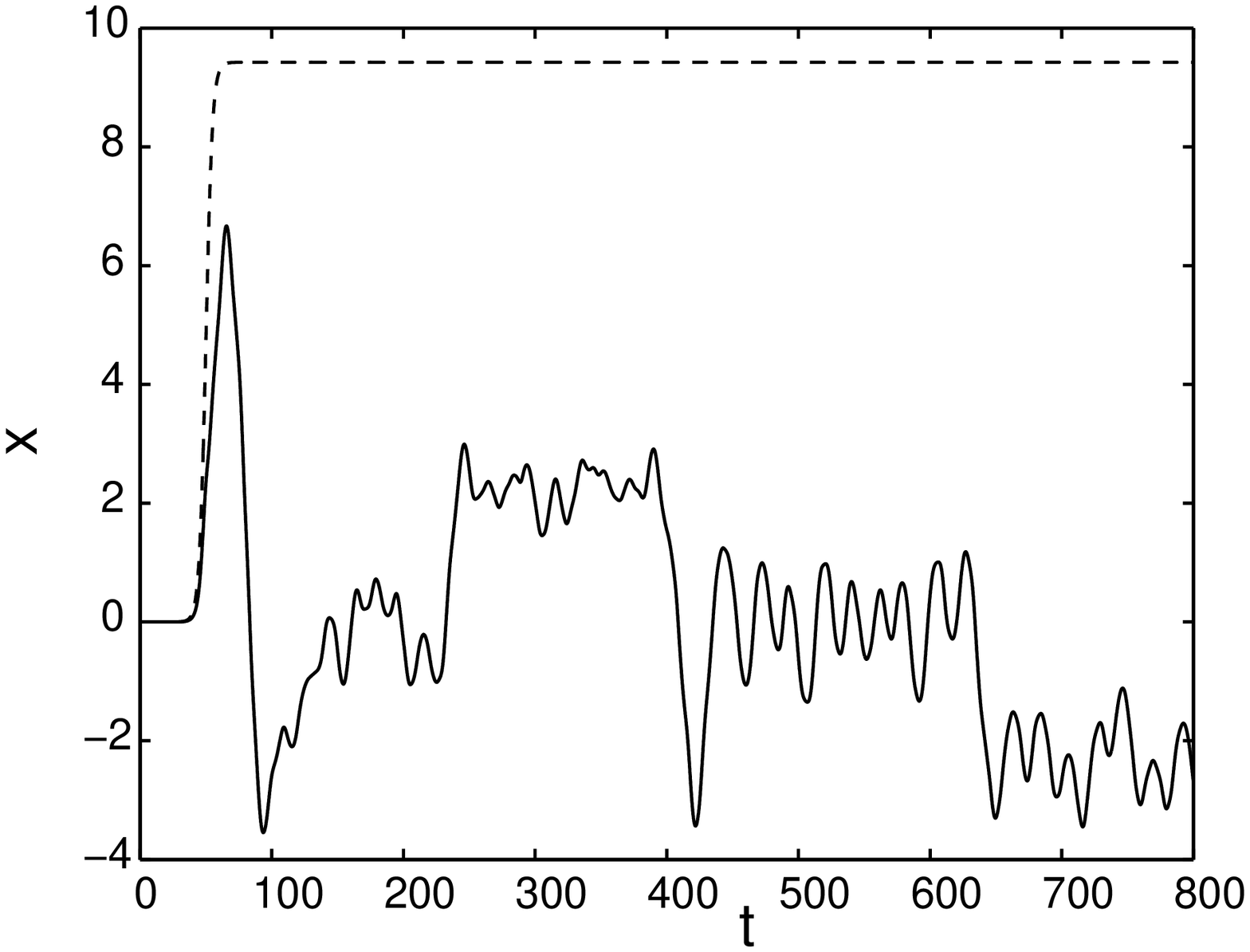}
\includegraphics[width=\wfig,height=\hfig,angle=0,clip]{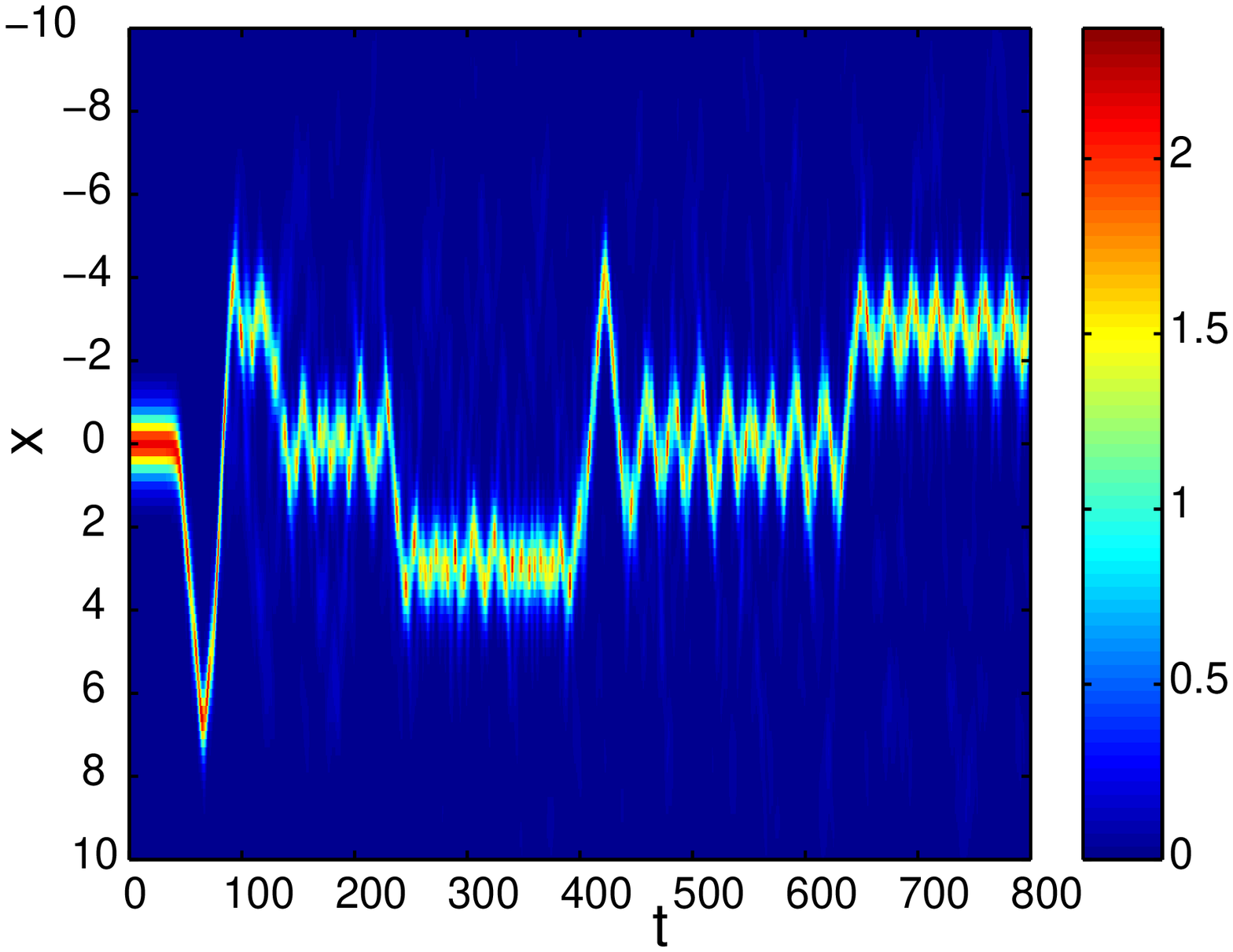}
}
\centerline{
\includegraphics[width=\wfig,height=\hfig,angle=0,clip]{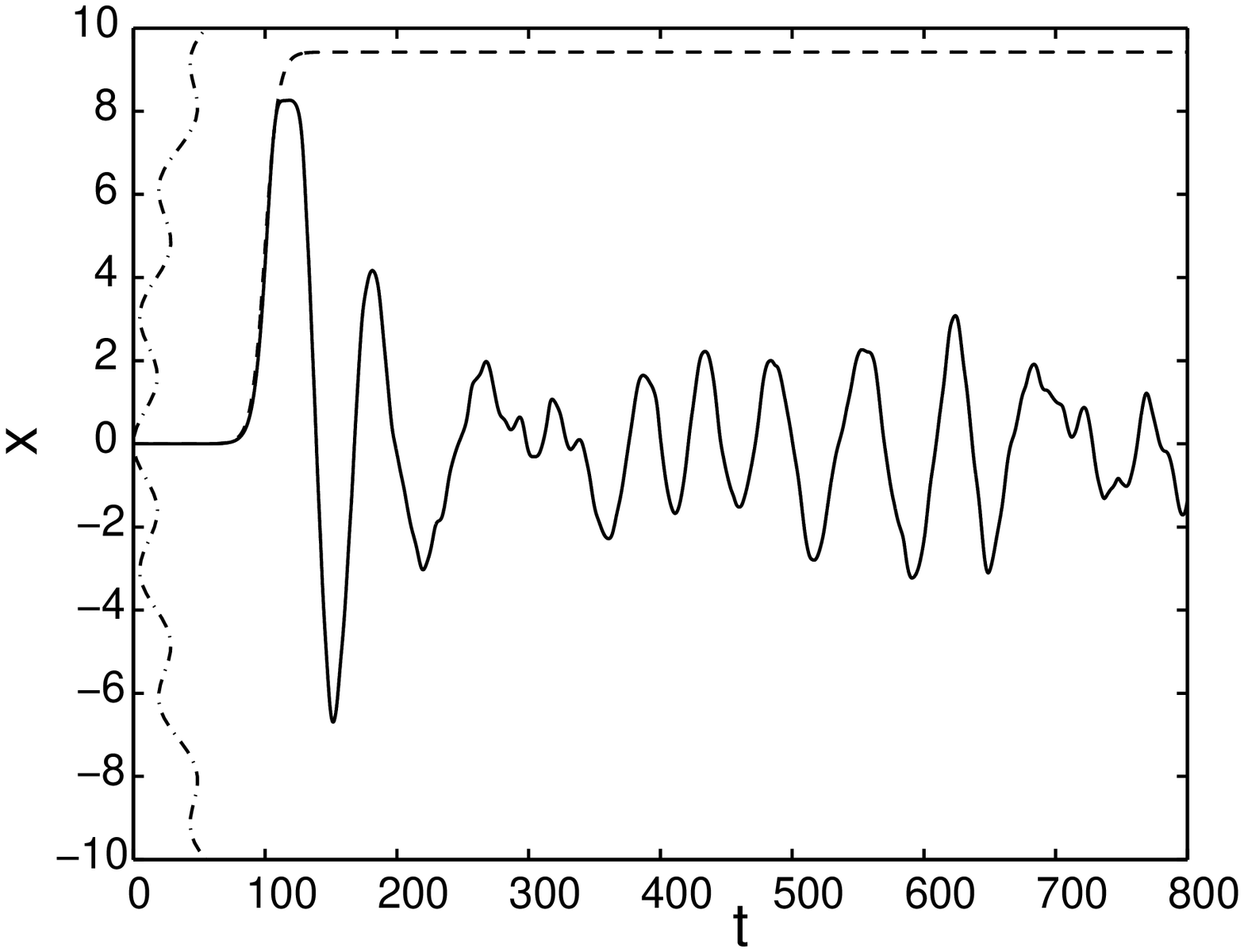}
\includegraphics[width=\wfig,height=\hfig,angle=0,clip]{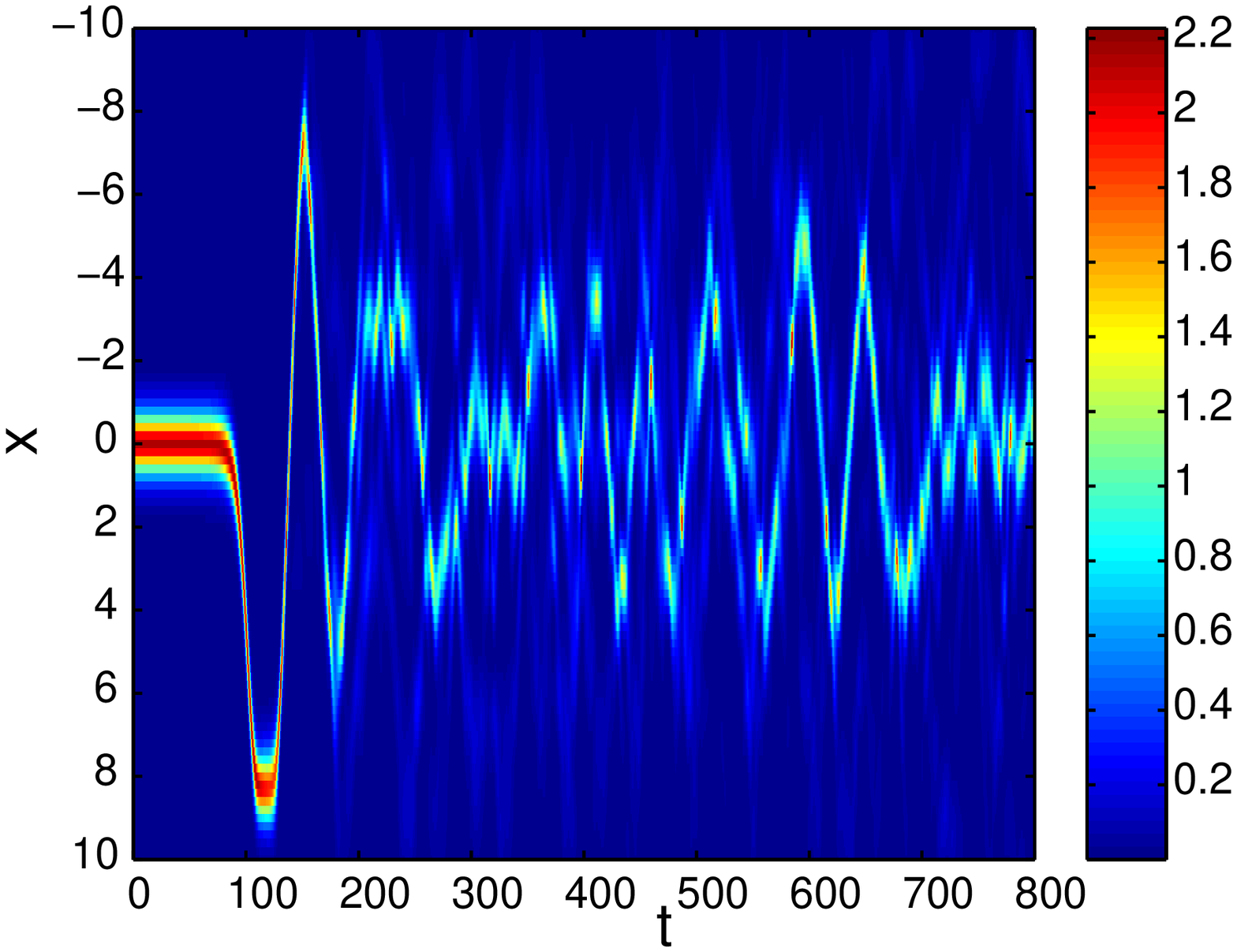}
} \caption{(color online) Panels have the same meaning as in Fig.\
\protect\ref{Fig1a}, but now for the case of a moving OL. The left
panel shows the soliton's center of mass by the solid line and the
motion of the OL's center by the dashed line. The potential
$V(x,t=0)$ is sketched by the dash-dotted line to illustrate the
structure and location of the potential wells. The right panel
again shows the space-time contour plot of $|u(x,t)|^{2}$. The top
set of the panels is generated with $t_{0}=50$ and $\protect\tau
=5$ in Eq.\ (\protect\ref{meq11}). The second set pertains to
$t_{0}=100$ and $\protect\tau =10$ (both have $V_{0}=0.25$). The
situation for a shallower well, with $V_{0}=0.17$, is shown in the
third and fourth panels. In all cases, $x_{\mathrm{ini}}=0$ and
$x_{\mathrm{fin}}=3\protect\pi $.} \label{Fig2}
\end{figure}
%%%%%%%%%%%%%%%%%%%%%%%%%%%%%%%%%%%%%%%%%%%%%%%%%%%%%%%%%%%%%%%%%%%%%%%%%%%%%

A new issue is whether one can capture the soliton performing such
oscillations by turning on the OL. Focusing, as previously, on the most
relevant case when the width of the soliton is comparable to the OL
wavelength, we display an example of the capture in Fig.\ \ref{Fig1a}. The
dashed and solid lines show, respectively, the harmonic oscillations in the
absence of the OL, and a numerical experiment, where, at the moment when the
soliton arrives at the turning point (it is $x=3\pi $ for this case, i.e.,
the third potential minimum), we abruptly turn on the OL, so that
\begin{equation}
V(x,t)=\frac{1}{2}\Omega ^{2}x^{2}+\frac{1}{2}V_{0}\left[ 1+\tanh \left(
\frac{t-t_{0}}{\tau }\right) \right] \sin ^{2}(kx).  \label{meq11}
\end{equation}Here $t_{0}$ and $\tau $ are constants controlling, respectively, the
switch-on time and duration of the process; in the simulations, we
use $t_{0}=31.7$ and $\tau =0.1$. We clearly observe that,
contrary to the large-amplitude oscillations of the soliton taking
place when the OL is absent, the soliton is now fully captured
(for very long times) by the potential minimum newly generated by
the optical trap.

Instead of being a means to snare for moving soliton, the OL
may also be used as a means of moving the soliton in a prescribed
way, i.e., as a ``robotic arm" depositing the soliton at a desired
location (see, e.g., \cite{hector}). This possibility is demonstrated (with varying levels of
success) in Fig.\ \ref{Fig2}. The top two sets of figures are
performed for a strong OL ($V_{0}=0.25$), while the bottom two are
used for a weaker OL potential (with $V_{0}=0.17$). In all the
cases the potential used is
\begin{equation}
V(x)=\frac{1}{2}\Omega ^{2}x^{2}+{V_{0}}\sin ^{2}\left( k(x-y(t))\right) ,
\end{equation}
where the position of the OL is translated according to
\begin{equation}
y(t)=x_{\mathrm{ini}}+\frac{1}{2}(x_{\mathrm{fin}}-x_{\mathrm{ini}})\left[
1+\tanh \left( \frac{t-t_{0}}{\tau }\right) \right] .
\end{equation}
Here $x_{\mathrm{ini}}$ and $x_{\mathrm{fin}}$ are, respectively, the
initial and final (target) positions of the soliton.
In the case under consideration, $x_{\mathrm{ini}}=0$ and
$x_{\mathrm{fin}}=3\pi $, i.e., the aim is to transport the MW
soliton from the central well to the third one, on the right of
the center. In the top set of the panels with $t_{0}=50$ and $\tau
=5$, we observe what happens if the motion of the potential center
is not sufficiently slow to adiabatically transport the soliton to
its final position. In particular, the soliton gets trapped in the
second well, without being able to reach its destination. However,
we observe that this difficulty can be overcome, if the transport
is applied with a sufficient degree of adiabaticity; see, e.g.,
the middle panel with $t_{0}=100$ and $\tau =10$, which succeeds
in delivering the soliton at the desired position. Notice that the
final position of the center of the OL is at $y=3\pi $, which is
different from the center of the third well of the effective
potential, around which the soliton will oscillate, upon arrival.
The theoretical prediction that was presented above (for
$V_{0}=0.25$) for this well is $x_{0}=9.0089$, while in the
simulations the soliton oscillates around $9.04$ in very good
agreement with the theory. The two lower sets of the panels are
meant to illustrate that adiabaticity is not the single condition
guaranteeing the efficient transport. The numerical experiments
are performed for a shallower potential where the relevant well
(to which the soliton is to be delivered) is near the threshold of
its existence. As a result, neither in the case with $\tau =5$
(the third set of panels), nor in the one with $\tau =10$, is the
OL successful in transporting the soliton at
the desired position. %%As a result of
%%the obtained potential energy (when escaping this well) the soliton performs
%%oscillations in the \textit{combined} potential.

\begin{figure}[t]
\centerline{
\includegraphics[width=\wfig,height=\hfig,angle=0,clip]{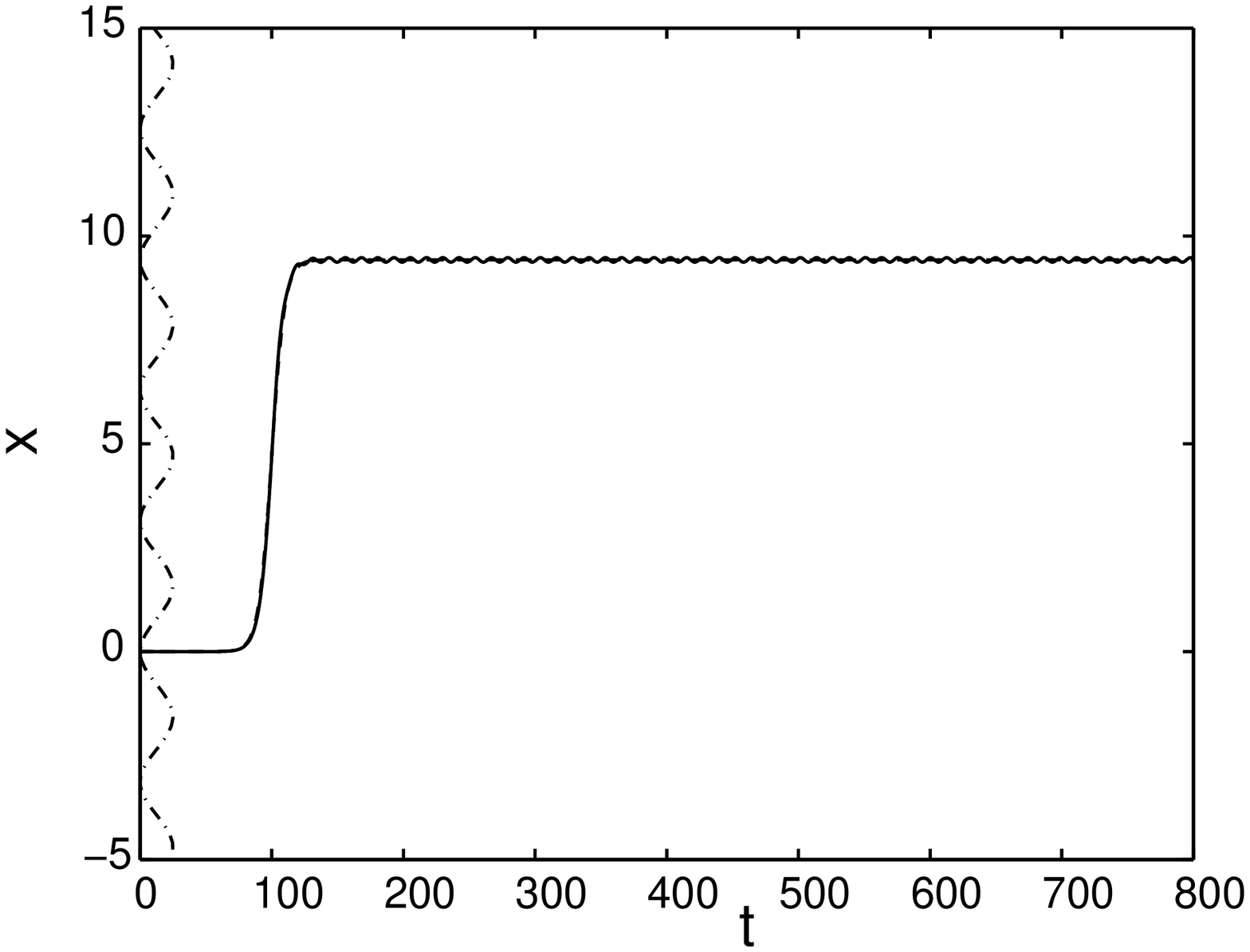}
\includegraphics[width=\wfig,height=\hfig,angle=0,clip]{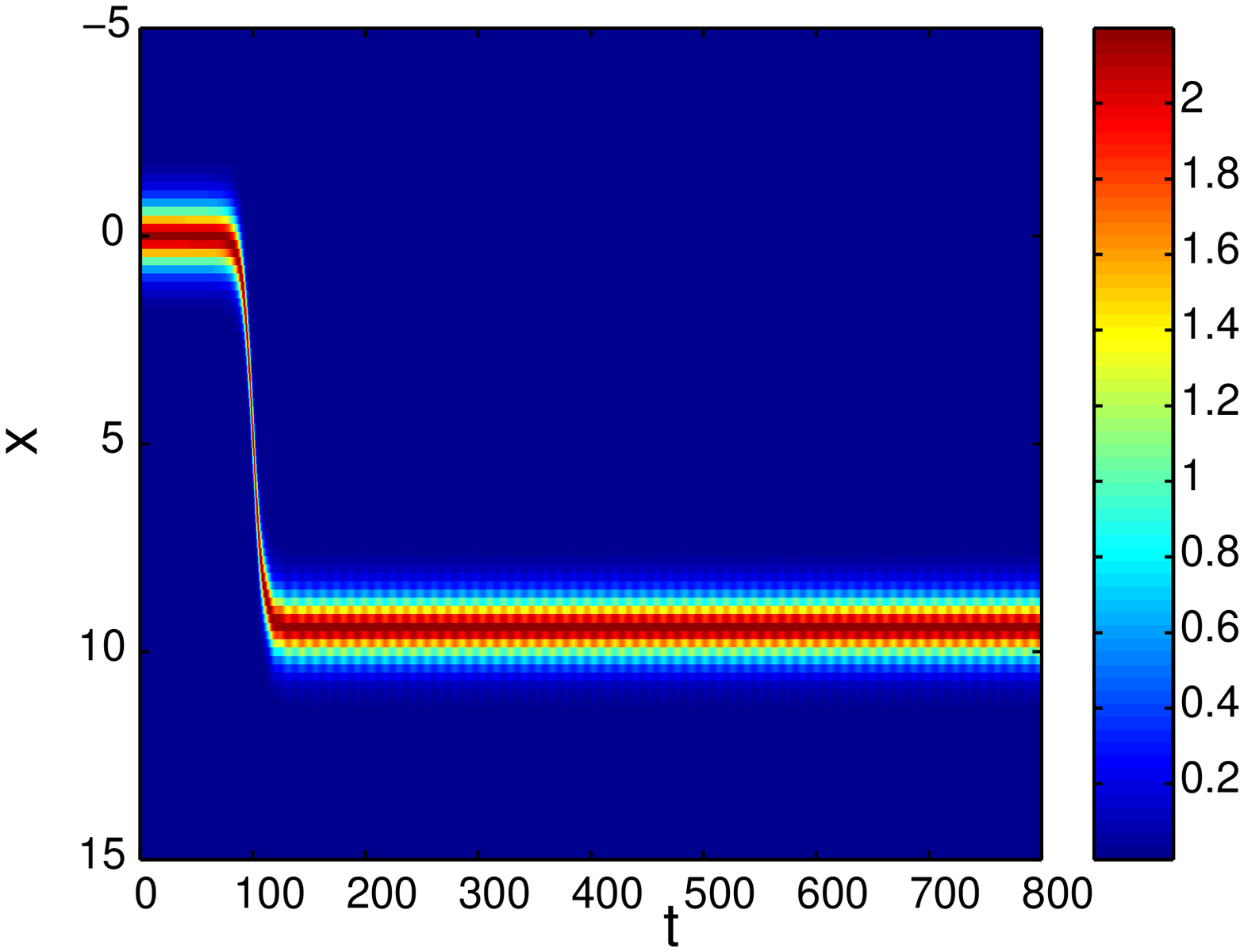}
}
\centerline{
\includegraphics[width=\wfig,height=\hfig,angle=0,clip]{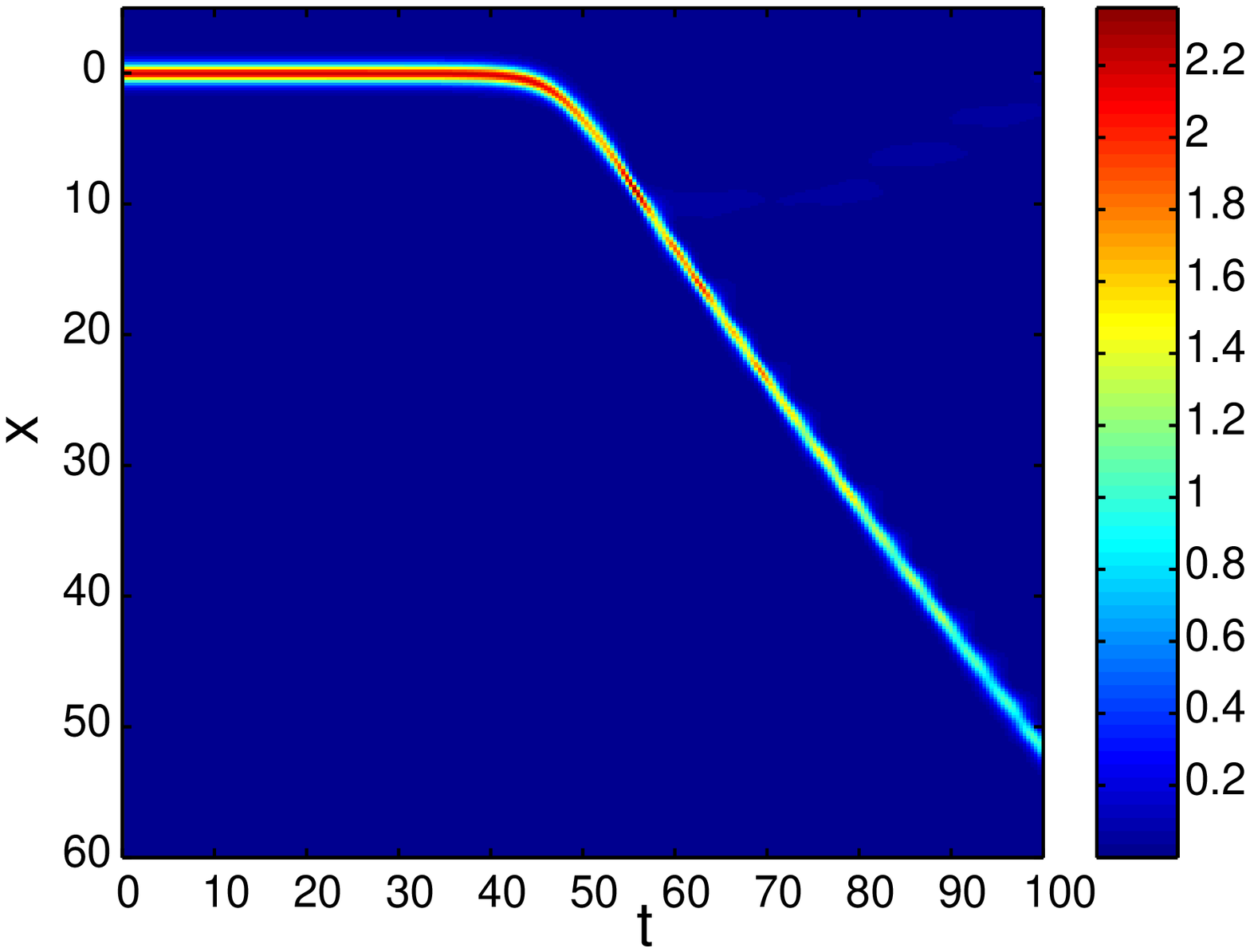}
} \caption{(color online) The same as the previous figure, but
with $\Omega =0$ (i.e., in the absence of the magnetic trap). For
$t_{0}=100$ and $\protect\tau =10$ (top panels) the soliton is
delivered to its final location of $x_{\mathrm{fin}}=3\protect\pi
$. However, the same is not true for $t_{0}=50$ and $\protect\tau
=5$ in the bottom panel, where the soliton fails to stop but
rather continues its motion, losing more and more of its power
through emission of radiation.} \label{Fig3}
\end{figure}

A similar numerical experiment in the absence of the magnetic trap
is shown in Fig.\ \ref{Fig3}. The top panels display the
successful transfer of the soliton by the OL of a form similar to
that in Eq.\ (\ref{meq11}), with $\Omega =0$ and $V_{0}=0.25$, for
$t_{0}=100$ and $\tau =10$. Notice that, in the present case, the
final positions of the OL's center and of the soliton coincide [as
the atomic potential and the effective potential for the soliton
have the same set of minima in this case, cf. Eq.\ (\ref{meq5})].
However, once again, the same experiment, if not performed with a
sufficient degree of adiabaticity (as in the bottom panel of Fig.\
\ref{Fig3}, with $t_{0}=50$ and $\tau =5$), is not successful in
depositing the soliton at the
prescribed location. %and instead of scattering from the
%the turning point as it did before with the parabolic trap,
Instead, in this case the soliton continues to move along the OL, emitting
radiation waves and decreasing its amplitude.

\begin{figure}[t]
\centerline{
\includegraphics[width=\wfig,height=\hfig,angle=0,clip]{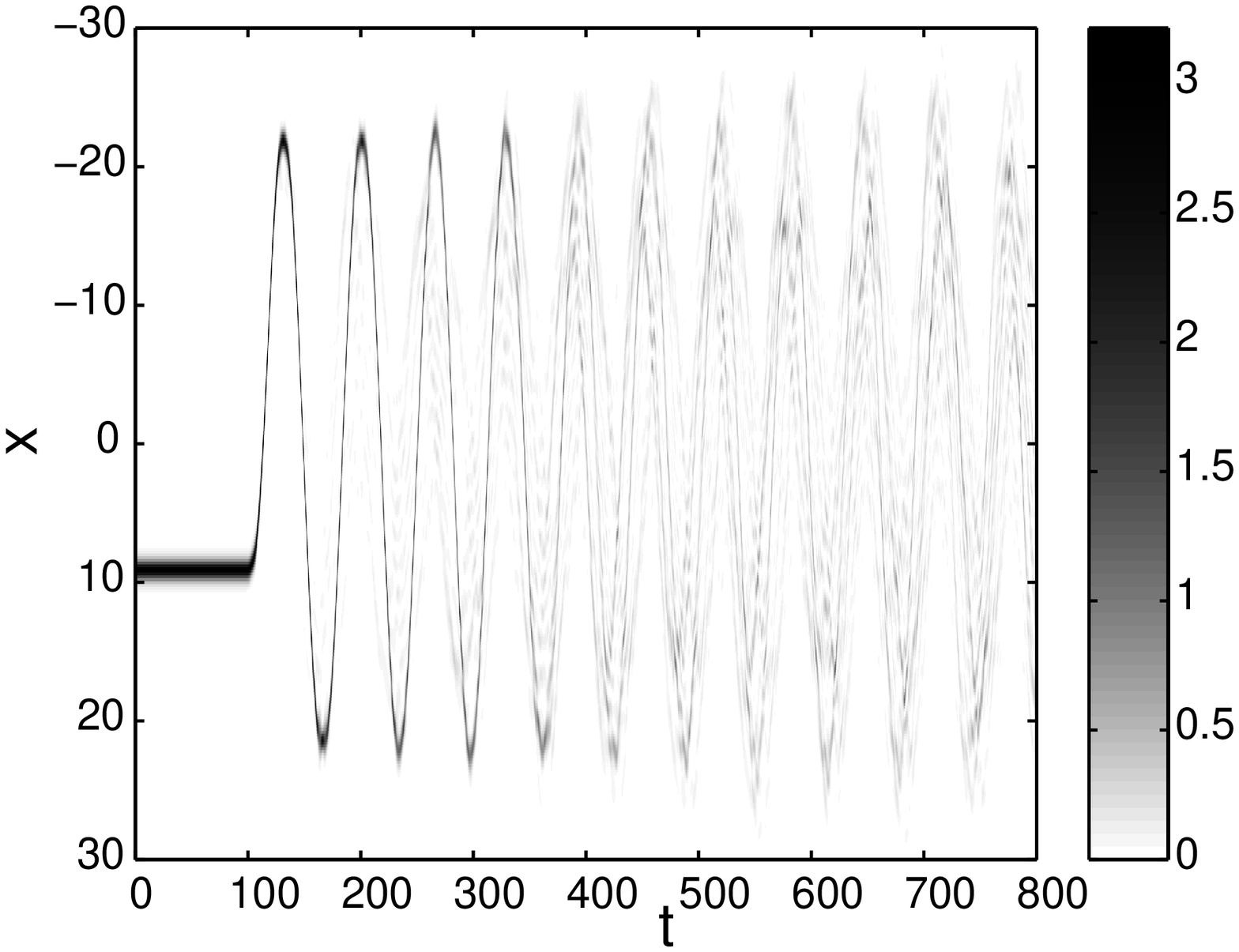}
\includegraphics[width=\wfig,height=\hfig,angle=0,clip]{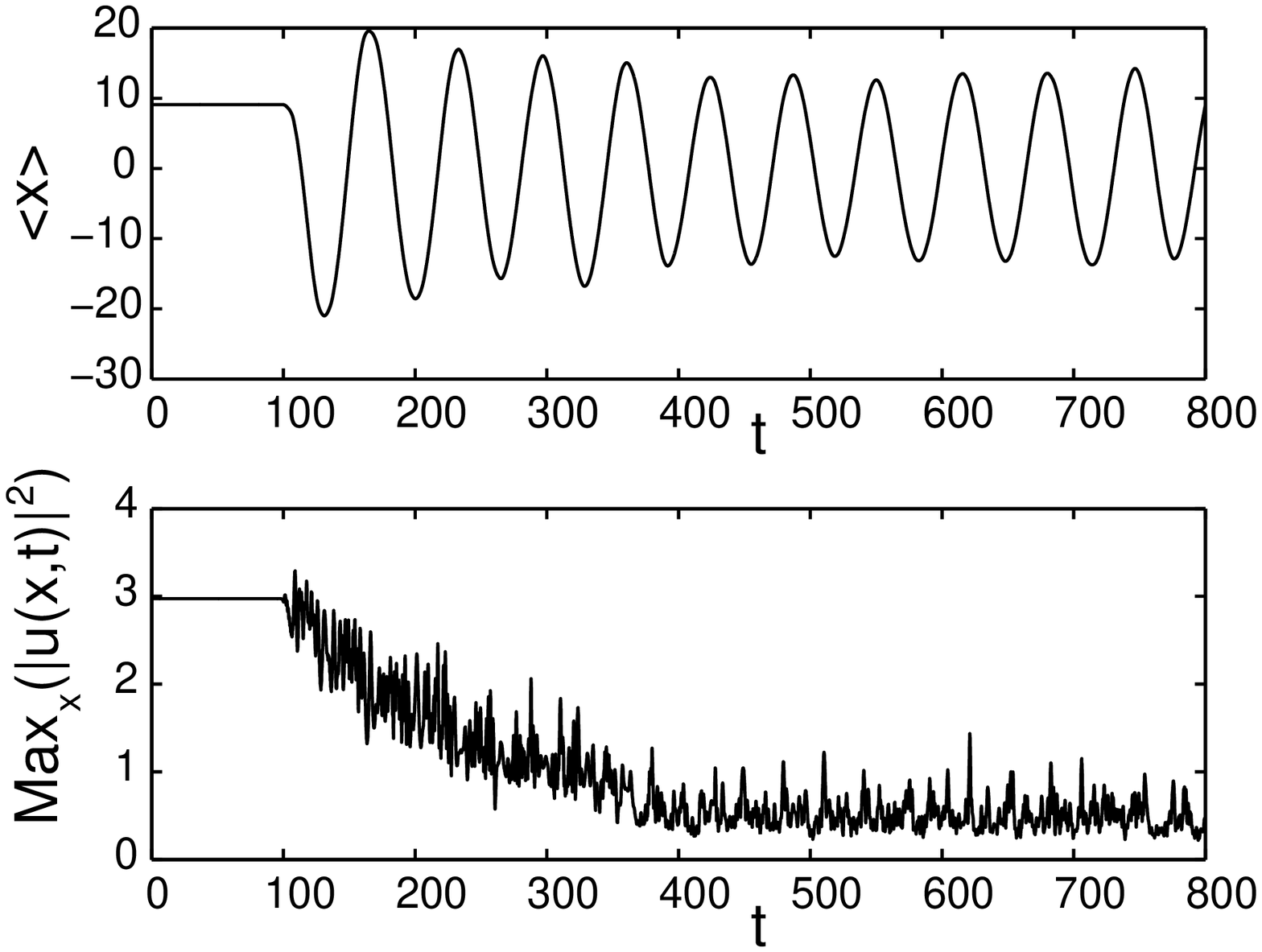}
}
\centerline{
\includegraphics[width=\wfig,height=\hfig,angle=0,clip]{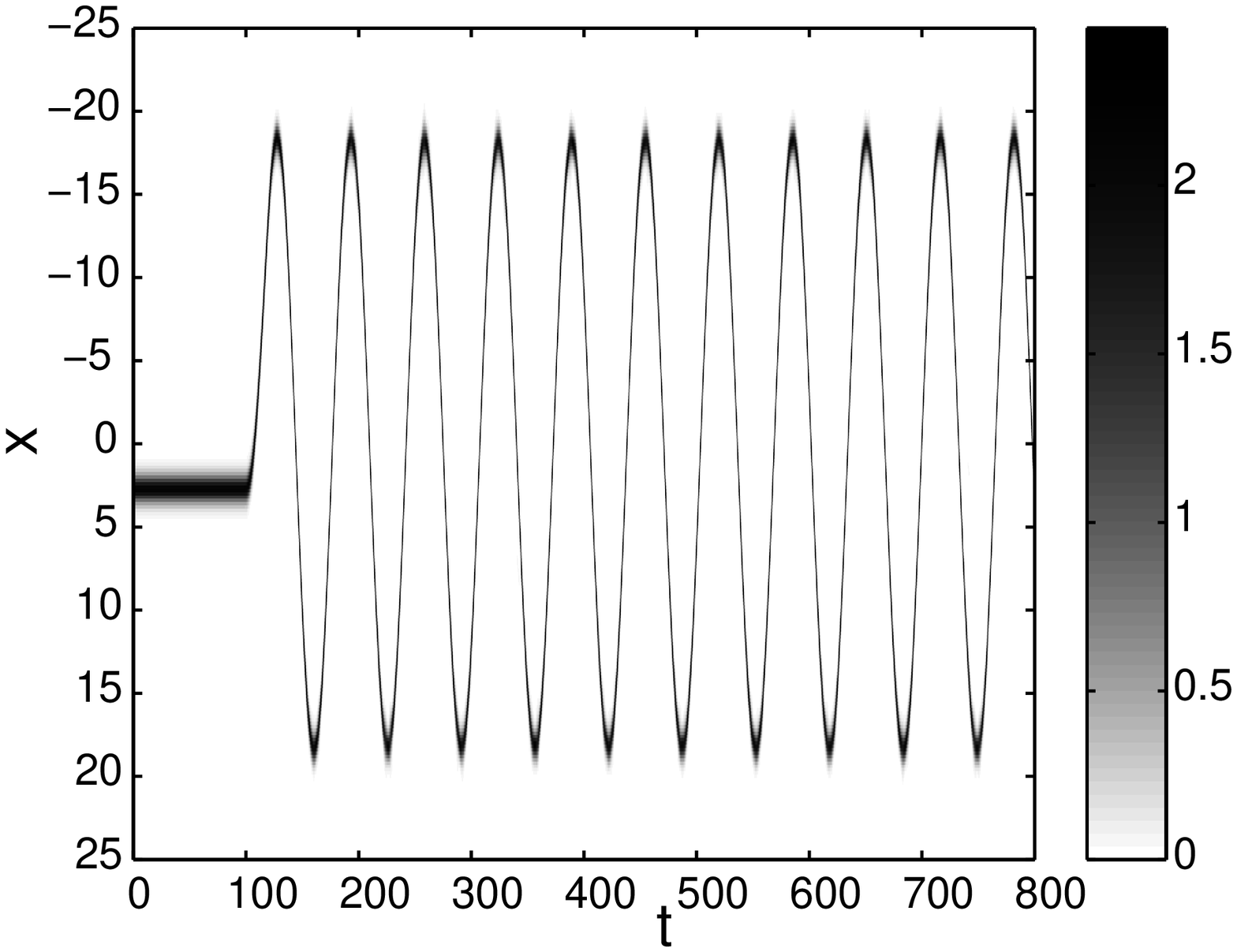}
\includegraphics[width=\wfig,height=\hfig,angle=0,clip]{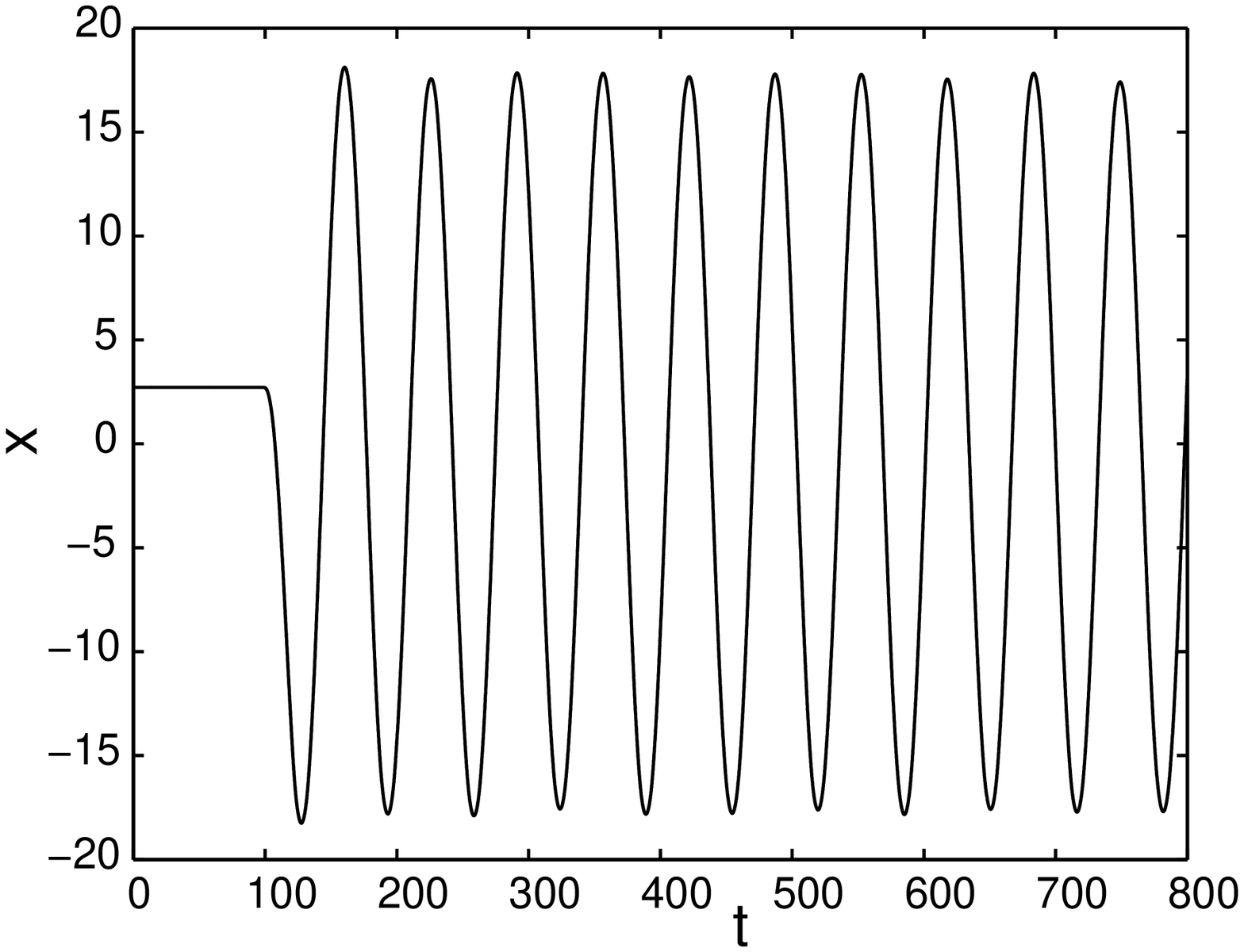}
}
\caption{Motion of the soliton induced by the linear ramp in Eq.\ (\protect
\ref{meq13}) with $t_{1}=100$ and $t_{2}=120$. The top panel shows the case
with strong radiation loss in a deep OL ($V_{0}=0.25$); notice apparent
friction in the motion of the soliton's center of mass in the top right
panel, and the corresponding loss of the soliton's norm in the panel below
it. On the contrary, in the case of the shallow OL, with $V_{0}=0.07$ (the
bottom panel), the moving soliton does not generate any 
visible radiation. In both
cases, $\protect\tau =1$ was used.}
\label{Fig4}
\end{figure}

To better illustrate the emission of radiation and its dependence on the
depth of the OL (since it is known that the emission is absent in the
parabolic potential without the OL ingredient), we have also performed the
following numerical experiment. We took the potential of the form
\begin{equation}
V(x)=\frac{1}{2}\Omega ^{2}x^{2}+V_{0}\sin ^{2}(kx)+\alpha (t)x,
\end{equation}with
\begin{equation}
\alpha (t)=0.1\times \frac{1}{2}\left[ \tanh \left(
\frac{t-t_{1}}{\tau }\right) -\tanh \left( \frac{t-t_{2}}{\tau
}\right) \right] .  \label{meq13}
\end{equation}In Eq.\ (\ref{meq13}), $t_{1}$ and $t_{2}$ are, respectively, the initial
and final moment of time, between which the linear ramp is applied
to accelerate the soliton to a finite propagation speed. We
display two such numerical simulations in Fig.\ \ref{Fig4}. The
first is performed in a deep OL, with $V_{0}=0.25$, taking
initially the soliton in the third well ($t_{1}=100$ and
$t_{2}=120$ were used). The second simulation was performed in a
shallow OL, with $V_{0}=0.07$, the soliton being initially taken
in the first well (the only one existing at such values of the
parameters). The top panels clearly show that the emission of
radiation leads to the gradual decay of the soliton's amplitude.
On the contrary, when the OL is weaker (in the bottom panels), the
soliton moves through it practically without radiation loss.

\section{Conclusion\label{sec4}}

We have examined a number of static and dynamic features of bright
matter-wave (MW) solitons in the presence of the magnetic trap and optical
lattice (OL). We used the perturbation theory to predict the existence and
stability of the MW solitons trapped in the combined potential. A sequence
of saddle-node bifurcations of the effective potential, which lead to
consecutive disappearance of the higher-well solitonic bound states with the
decrease of the OL strength was predicted, through the disappearance of the
potential wells in the effective potential.

Having identified the stability characteristics of the different wells
analytically, and verified it numerically, we then explored a possibility to
use the OL as a tool to manipulate the soliton. We were able to stop the
soliton at a prescribed location by turning on the OL, in an appropriate
fashion. We have also found the adiabaticity condition necessary to secure
the transfer of the soliton by a moving OL (with and without the magnetic
trap). Finally, we have shown the absence of any visible emission of
radiation from the soliton moving across a weak OL; however, the soliton
loses a large fraction of its norm, moving through a stronger lattice.

Given the recent prediction of solitons and vortices in multi-dimensional OL
potentials \cite{md} (for recent experimental work on a similar topic in
nonlinear optics, see Refs.\ \cite{neshev,moti} and references therein), it
would be of particular interest to implement similar dragging and
manipulation of solitons in higher dimensions. The consideration of this
case is currently in progress.

This work was partially supported by NSF-DMS-0204585, NSF-CAREER, and the
Eppley Foundation for Research (PGK); the Israel Science Foundation grant
No.\ 8006/03 (BAM); and the San Diego State University Foundation (RCG). PGK
also gratefully acknowledges the hospitality of the Center for Nonlinear
Studies of the Los Alamos National Laboratory. Work at Los Alamos is
supported by the US DoE.

\end{document}